\documentstyle[11pt]{article}
\def\a{\alpha}\def\b{\beta}\def\g{\gamma}\def\d{\delta}\def\e{\epsilon}

\def\k{\kappa}\def\l{\lambda}\def\L{\Lambda}\def\s{\sigma}\def\S{\Sigma}
\def\Th{\Theta}\def\om{\omega}\def\G{\Gamma}
\newcommand{\p}[1]{(\ref{#1})}
\newcommand{\plabel}[1]{\label{#1} 
}
\begin{document}
\renewcommand{\thefootnote}{\fnsymbol{footnote}}
\setcounter{page}0
\thispagestyle{empty}
\def\theequation{\thesection.\arabic{equation}}

\begin{flushright}
{\bf FTUV/99-07, IFIC/99-07 \\
TUW-99-06 \\
HUB-EP-99/15\\
hep-th/9904109} \\
 April 14, 1999
\end{flushright}

\bigskip

\begin{center}
{\Large \bf 
Superparticle Models with 
Tensorial Central Charges}

\bigskip

{\large \bf 
Igor Bandos$^{2)}$\footnote{ Lise Meitner Fellow.
On leave of absence from  Institute for Theoretical Physics,
NSC Kharkov Institute of Physics and Technology,
310108 Kharkov, Ukraine}, 
Jerzy Lukierski$^{1),3)}$\footnote{Supported in part by {\bf KBN} grant
{\bf 2P03B13012}} and 
Dmitri Sorokin}$^{4)}$\footnote{A. von Humboldt Fellow.
On leave of absence from  Institute for Theoretical Physics,
NSC Kharkov Institute of Physics and Technology,
310108 Kharkov, Ukraine},

\bigskip
$^{1)}$ {\it 
Dpto. de Fisica Teorica and IFIC (CSIC), 
Facultad de Fisica \\
E-46100-Burjassot (Valencia), 
 Spain}\\
$^{2)}$ {\it Institute for Theoretical Physics, 
Technical University of Vienna\\
Wiedner Haupstrasse 8-10, 
A-1040 Wien, Austria}\\
{\bf e-mail: bandos@tph32.tuwien.ac.at}
 \\
$^{3)}${\it Institute for Theoretical Physics 
University of Wroclaw, \\
50-204 Wroclaw, Poland}
\\
{\bf e-mail: lukier@proton.ift.uni.wroc.pl}
\\
$^{4)}${\it  Humboldt-Universit\"at zu Berlin\\
Mathematisch-Naturwissenshaftliche Fakultat, 
Institut f\"ur Physik, \\
Invalidenstrasse 110, D-10115 Berlin, Germany\\
{\bf e-mail: sorokin@physik.hu-berlin.de}
}

\end{center}


\bigskip

\begin{abstract}

A generalization of the Ferber--Shirafuji formulation of 
superparticle mechanics is considered. The generalized model
describes the dynamics of a superparticle in a superspace extended
by tensorial central charge coordinates and commuting twistor--like 
spinor variables.
The $D=4$ model contains a continuous real parameter $a\geq 0$ and at
$a=0$ reduces to the $SU(2,2|1)$ supertwistor Ferber--Shirafuji model, 
while at
$a=1$ one gets an $OSp(1|8)$ supertwistor
model of ref. \cite{BL} which describes BPS states with all but one
unbroken target space supersymmetries. 
When  $0<a<1$ the model admits
an $OSp(2|8)$ supertwistor description, and 
when $a>1$ the supertwistor group becomes $OSp(1,1|8)$.
We quantize the model and find that its quantum spectrum consists of 
massless states of an arbitrary (half)integer helicity. The independent
discrete central charge coordinate describes the helicity spectrum.

We also outline the  generalization
of the 
$a=1$ model to higher space--time dimensions and demonstrate that in  
$D=3, 4, 6$ and $10$, where the quantum states are 
massless, the  
extra degrees of
freedom (with respect to those of the standard superparticle) 
parametrize compact manifolds. These compact manifolds can be associated
with higher--dimensional helicity states.
In particular, in $D=10$ the additional ``helicity" manifold is
isomorphic
to the sphere $S^7$.

\end{abstract}

\newpage

\renewcommand{\thefootnote}{\arabic{footnote}}
\setcounter{footnote}{0}

\section{Introduction}
\setcounter{equation}{0}

In a recent paper \cite{BL} two of the present authors proposed a new
superparticle model with tensorial central charges \cite{HP}--\cite{RS} 
and 
auxiliary fundamental spinor variables. An interesting peculiar feature
of this model is that it describes a superparticle whose presence breaks
only one of target--space supersymmetries. In all previously known cases
superparticles and (in general) superbranes break half or more 
supersymmetries of a target superspace vacuum. 

As we shall show in this paper
the model of \cite{BL} describes an infinite tower of 
massless particles of arbitrary
(half)integer helicities.

The model can be
regarded as an extension of a Ferber--Shirafuji formulation
\cite{F78,S83} of $D=4$
 superparticle mechanics. In the framework of
the $N=1$ Ferber--Shirafuji model one performs,
at the
classical level, the twistor transform \cite{F78,S83} from the $N=1$, $D=4$
superspace
description of massless superfields  to their
description 
in terms of supertwistors forming a fundamental representation of
a superconformal group $SU(2,2|1)$.

In an analogous way the superparticle model of \cite{BL} admits the 
description in terms of $OSp(1|8)$ supertwistors \cite{Lstw,8}.

The supergroups $SU(2,2|1)$ and $OSp(1|8)$ are not subgroups of each
other,
but they are different subgroups of the supergroup $OSp(2|8)$. Hence,
one can assume that the Ferber--Shirafuji model and the model \cite{BL}
are different reductions of an $OSp(2|8)$ supertwistor model.

In this paper we construct such a generic $N=1$, $D=4$ superparticle
action
which depends on a numerical non--negative real parameter $a$.
When the value of $a$ varies within the interval $0<a<1$ the model
admits
an $OSp(2|8)$ supertwistor description, while $a=0$ and $a=1$ are two
critical points. At $a=0$ the model reduces to the Ferber--Shirafuji
superparticle. And at $a=1$ one arrives at the $OSp(1|8)$ supertwistor
model of ref. \cite{BL}. 

For all values of $a$ except for $a=1$ the superparticle breaks half of
the
target--space supersymmetries, while at $a=1$ only one supersymmetry is
broken.

When $a>1$ the supertwistor group becomes $OSp(1,1|8)$ which contains
a noncompact group $SO(1,1)$ as a subgroup instead of $SO(2)$ in the
case of
$OSp(2|8)$.
 
The (super)twistor formulation of relativistic (super)particle dynamics
is useful in many aspects. Let us recall that, since the relativistic
(super)particle is a constrained dynamical system not all its dynamical
variables are independent. By performing (super)twistor transform
we deal directly with independent physical degrees of freedom
of the (super)particle in a covariant way. This, for instance,
simplifies
the quantization procedure and the analysis of the spectrum of quantum
states of the model.

We perform the quantization of the generic superparticle model for
arbitrary values of the parameter $a$ and find that in $D=4$ 
first--quantized states of the superparticle form an infinite tower of 
massless states of a (half)integer helicity. 
We thus demonstrate that an extra dynamical
(central charge) coordinate in the model under consideration has the
physical meaning of a spin variable. This allows one to admit that
the model considered might be related to the higher--spin field theory
of Vasiliev \cite{Vasiliev} (see also relevant papers \cite{rel}).

We first quantize the superparticle in the supertwistor formulation
where
the quantization is almost straightforward since at $a\not =0$ the 
supertwistor model is unconstrained and at $a=0$ there is only one 
first--class constraint. 

We then quantize the model in the $N=1$, $D=4$ superspace extended with 
the tensorial central charge coordinates, and show that the resulting
spectrum of the quantum physical states coincides with that of the 
supertwistor formulation.

Since in this formulation the model contains second
class constraints 
our main tool in carrying out the quantization procedure 
will be
the extension of the model in such a way that all the
constraints of the initial model 
become first class constraints. This method, which
can be traced back to the papers by 
Faddeev \& Shatashvili \cite{9,10}, 
Batalin, Fradkin \& Fradkina \cite{12,13} and 
Egorian \& Manvelian \cite{11}  
has already been applied 
to the quantization of `standard' massless
superparticles by Moshe \cite{14} and Eisenberg \& 
Solomon \cite{15,ES,17}. 
The
main advantage of this method  is that it allows one to avoid problems
with covariant splitting fermionic
constraints into first and second class ones. 
The initial formulation of the 
 model is recovered when we 
gauge fix additional 
gauge symmetries (associated with new first--class constraints)
by putting the conversion variables to zero. 
In its nature the conversion method is
related to an old Stueckelberg formalism \cite{18} which extends the
theory
of massive vector fields with  
an auxiliary scalar gauge degree of freedom. 

The quantization of the model at $a=1$ has additional peculiarities. 
In this case superparticle dynamics is subject to only one second--class
constraint, which is quite unusual. Dynamical systems with the odd
number
of fermionic second class constraints are rather rare. One of few known
examples is a superparticle in $D=2$ superspace with a single chiral 
fermion direction \cite{Sorokin87}. So the quantization of such systems
is an 
interesting exercise by itself which requires one to deal with a single
Clifford--like variable. In the case under consideration we shall use
an auxiliary Clifford variable of Grassmann--odd parity to convert the
 single 
second class fermionic constraint into  
 the one of 
  a first class. Further we   present
two  methods for quantizing the model with the single Clifford
variable,
both producing the same spectrum of first--quantized physical
states.

The paper is organized as follows.

In Section 2 we consider the one-parameter family of actions 
describing the  generalized $D=4$ superparticle models labelled 
by the real positive parameter $a\ge 0$ where the case $a=0$  
corresponds to the Ferber--Shirafuji model, 
while at $a=1$ the action describes the superparticle model of
\cite{BL}. We demonstrate that in the target space with four 
supersymmetries the $a\not=1$ models possess two fermionic 
$\kappa$-symmetries and, hence, corresponding superparticle 
configurations preserve 1/2 of the supersymmetries,
while the $a=1$ generalized superparticle has three $\kappa$--symmetries 
and, hence, preserves $3\over 4$ of the supersymmetries.   
We also find that the $U(1)$ symmetry is inherent to the $a=0$ case
only.

In Section 3 we describe the transform to a supertwistor form of the
action. 
We show that $0<a<1$ models are described by a
free $OSp(2|8)$ supertwistor action and thus link 
the Ferber-Shirafuji $SU(2,2|1)$ supertwistor model 
and the free $OSp(1|8)$ supertwistor model 
of \cite{BL}. 
The model with  $a>1$ is transformed into a free $OSp(1,1|8)$
supertwistor 
action. 
We perform 
the quantization of the 
supertwistor models and find that 
the `supertwistor' wave function describes an infinite tower of 
short supersymmetric multiplets of massless fields of all possible 
helicities. 

In Section 4 we extend the initial
phase space of the superparticle model with auxiliary variables and
perform the conversion of the initial set of first and
second class constraints
into the first--class constraints generating new gauge
symmetries.  We then carry out the quantization of the
extended model and find that the dependence of the wave functions
on Grassmann--odd conversion variables is inessential and can be
ignored. 
We show that 
the wave function of the first--quantized model
of Section 2 can be identified with the supertwistor wave
function of Section 3 if Cartan--Penrose twistor formulae 
relating superspace and supertwistor coordinates are imposed. 
Thus, we find that the infinite spectrum of the first--quantized states
of
the superparticle consists of massless fields of an arbitrary
(half)integer helicity.

In Section 5 we consider a  multidimensional generalization of the 
$a=1$ model and its quantization. It appears that after
quantization the superwave function depends on only one Grassmann
variable, and all other fermionic degrees of freedom can be
eliminated by $N-1$  $\kappa$--transformations, where $N$ is the
total number of supersymmetries. Thus, the corresponding superparticle 
configuration preserves the $(N-1)/N$ fraction of 
target--space supersymmetry.

In the Appendix we analyze in detail the quantization of a
supersymmetric system with one real Grassmann variable, which
after quantization becomes a single Clifford variable.

\section{$D=4$ model with fundamental spinor and tensorial central 
charge coordinates}
\setcounter{equation}{0}

Let us consider the following $D=4$ superparticle action 
\begin{equation}\plabel{ac(a)}
S = \int d \tau
\left( \lambda_{A}\bar{\lambda}_{\dot{B}} \Pi_\tau^{A\dot{B}} +
a \lambda_{A}\lambda_{B}  \Pi_\tau^{AB}
+ \bar a \bar{\lambda}_{\dot{A}} \bar{\lambda}_{\dot{B}}
\, \Pi_\tau^{\dot{A}\dot{B}} \right)\, ,
\end{equation}
where 
\begin{equation}\plabel{vielbeine}
\begin{array}{l}
\displaystyle
\Pi^{A\dot{B}} \equiv  d\tau \Pi_\tau^{A\dot{B}}
= dx^{A\dot{B}}
+ i \left( d\Theta^{A} \bar{\Theta}^{\dot{B}}
- \Theta^{A} d\bar{\Theta}^{\dot{B}}\right) \, ,
\\ \nonumber
\displaystyle
\Pi^{AB} \equiv d\tau \Pi_\tau^{AB} =
d{y}^{AB} - ~i~ \Theta^{(A}~d{\Theta}^{B)}\, ,
\\  
\displaystyle
\bar{\Pi}^{\dot{A}\dot{B}}
\equiv d\tau \bar{\Pi}_\tau^{\dot{A}\dot{B}}
= d{\bar{y}}^{\dot{A}\dot{B}}
- ~i ~\bar{\Theta}^{(\dot{A}} ~d{\bar{\Theta}}\,{}^{\dot{B})}\, ,
\end{array}
\end{equation}
 $A, B=1,2$, $\dot{A},\dot{B} =1,2$
are Weyl spinor indices, and
the spin--tensors $x^{A\dot{B}}$ and ${y}^{AB}$ are related to $D=4$
vector
coordinates $x^m$ and antisymmetric tensorial coordinates $y^{mn}$
through the Pauli matrices
\begin{equation}\label{vt}
x^{A\dot{B}}=x^m\sigma_m^{A\dot{B}}, \qquad {y}^{AB}=
{1 \over 2} y^{mn}(\sigma_{[m}\tilde\sigma_{n]})^{A{B}} 
= ( \bar{y}^{\dot{A}\dot{B}})^*
\end{equation}
 $a$ is a numerical parameter which, without the loss of generality,
can be taken to be real and positive definite $a=\bar a \in [0, \infty
)$.
  Indeed, 
 if $a$ is complex its phase can always be absorbed by  
the bosonic spinor $\l_A$ redefined in an appropriate way 
 ($\l_A \rightarrow (\bar{a}/|a|)^{1/2} \l_A$).

The action \p{ac(a)} describes a superparticle propagating in the
extended 
superspace 
\begin{equation}\plabel{superspace}
M^{(4+6|4)} = \{ Y^M \} \equiv \{ (x^{A\dot{A}}, y^{AB},
\bar{y}^{\dot{A}\dot{B}}; \Theta^{A}, \bar{\Theta}^{\dot{A}})\}
\end{equation} 
{\sl with tensorial central charge coordinates}
 $y^{AB}, \bar{y}^{\dot{A}\dot{B}}$ 
\footnote{For previous consideration of different 
models of superparticles and 
p-branes in superspaces with tensorial central charges see 
\cite{ES,Cz,RS}.}. 
The configuration 
space of the system 
\begin{equation}\plabel{confsuperspace}
{\cal M}^{(4+6+4|4)} = \{ q^{{\cal M}} \} \equiv
\{  (Y^M; \l^A, \bar{\l}^{\dot{A}}) \}
=\{ (x^{A\dot{A}}, y^{AB},
\bar{y}^{\dot{A}\dot{B}};  \l^A, \bar{\l}^{\dot{A}};
\Theta^{A}, \bar{\Theta}^{\dot{A}})\}
\end{equation}
contains in addition  four  bosonic spinor coordinates 
$\l^A, \bar{\l}^{\dot{A}}$. 

The presence of the parameter $a$ in the action \p{ac(a)} reflects the
 property
that each of its three terms is  separately invariant under 
global supersymmetry transformations acting on $M^{(4+6|4)}$ as follows
$$
\d \Theta^A = \e^A, \qquad  \d \bar{\Theta}^{\dot{B}}=
\bar{\e}^{\dot{B}}, 
$$
\begin{equation}\plabel{susy}
\d x^{A\dot{B}} = i \e^A \bar{\Theta}^{\dot{B}} 
- i \Theta^A \bar{\e}^{\dot{B}}, \qquad 
\d {y}^{AB}= ~i~ \e^{(A}~{\Theta}^{B)}, \qquad 
d{\bar{y}}^{\dot{A}\dot{B}}=~i ~\bar{\e}^{(\dot{A}}
~{\bar{\Theta}}\,{}^{\dot{B})}, \qquad 
\end{equation}
$$
\d\l_A =0, \qquad \d \bar{\l}_{\dot{A}}=0.
$$

The generators of the transformations \p{susy} 
$$ \d Y^M = i (\e^AQ_A +\bar{Q}_{\dot{A}}\bar{\e}^{\dot{A}}) Y^M $$
satisfy the supersymmetry algebra with central charges 
\begin{equation}\plabel{alg}
\{ Q_A, Q_B\} = Z_{AB}, \qquad  
\{ Q_A, {\bar{Q}}_{\dot{B}} \} = - 2 P_{A\dot{B}}, \qquad  
\{\bar{Q}_{\dot{A}},  \bar{Q}_{\dot{B}}\} = \bar{Z}_{\dot{A}\dot{B}},
\qquad  
\end{equation}
and all other commutators of the generators vanish.

The superalgebra \p{alg} has the following realization in the
superspace  \p{superspace}

\begin{equation}\plabel{genera}
Q_A = i \partial_A - \partial_{A\dot{B}} \bar{\Theta}^{\dot{B}} -
{1 \over 2} \Theta^B \partial_{AB}, \qquad 
 \bar{ Q}_{\dot{A}} = - i \bar{\partial}_{\dot{A}} 
+ \partial_{\dot{A}{B}} \Theta^B + {1 \over 2}
\bar{\partial}_{\dot{A}\dot{B}}
\bar{\Theta}^{\dot{B}}, \qquad  
\end{equation}
\begin{equation}\plabel{gener}
P_{A\dot{B}} = - i \partial_{A\dot{B}}, \qquad 
Z_{AB} = - {i}\partial_{AB} , \qquad
\bar{Z}_{\dot{A}\dot{B}}= - {i} \bar{\partial}_{\dot{A}\dot{B}}
\end{equation}
$$
\partial_{A\dot{B}}\equiv { \partial \over \partial x^{A\dot{B}}}, 
\qquad 
\partial_{AB}\equiv { \partial \over \partial y^{AB}}, 
\qquad 
\bar{\partial}_{\dot{A}\dot{B}}\equiv { \partial \over \partial 
\bar{y}^{\dot{A}\dot{B}}}, 
$$
$$
\partial_{A} \equiv  { \partial \over \partial 
\Theta^A }, \qquad  \bar{\partial}_{\dot{A}}\equiv 
 { \partial \over \partial\bar{\Theta}^{\dot{A}}}.
$$

If $a=0$ the action \p{ac(a)} reduces to the Ferber--Shirafuji action
\cite{F78,S83},
and if $a=1$ the action becomes the one considered in  \cite{BL}.
We shall see that $a=0$ and $a=1$ are `critical' 
 values  of the parameter
where symmetries of the action \p{ac(a)} as well as
  the physical content
of
the model are  modified.

\subsection{Critical points $a=0$ and $a=1$} 
In order to analyze symmetry properties of the action \p{ac(a)} at
different
values 
of $a$ we  consider the general variation of 
\p{ac(a)} which (modulo boundary terms) has the form 
\begin{equation}\plabel{dac(a)}
\delta S = \int 
\left[ 
\d\lambda_{A}
\Big(\bar{\lambda}_{\dot{B}} \Pi^{A\dot{B}} +
2a \lambda_{B}  \Pi^{AB}\Big)
+ \d \bar{\lambda}_{\dot{A}} \Big(\Pi^{B\dot{A}}\lambda_{B} +
 2{a}  \bar{\lambda}_{\dot{B}} \Pi^{\dot{A}\dot{B}}
\Big) \right]\, - 
\end{equation}
$$
- \int 
\left[ d \Big(\lambda_{A}  \bar{\lambda}_{\dot{B}} \Big) i_\d
\Pi^{A\dot{B}} 
+ a ~d \Big(\lambda_{A} \lambda_{B}\Big) i_\d \Pi^{AB} + 
 {a} ~ d \Big(\bar{\lambda}_{\dot{A}}\bar{\lambda}_{\dot{B}}\Big)
 i_\d\Pi^{\dot{A}\dot{B}} \right]\, + 
$$
$$
+ \int 
\left[2i \Big( d\bar{\Theta}^{\dot{B}}
 \bar{\lambda}_{\dot{B}} + a d{\Theta}^{{B}} \l_B \Big) \d{\Theta}^{A}
\l_A 
+ 2i \Big(d{\Theta}^{{B}} \l_B +{a} d\bar{\Theta}^{\dot{B}}
 \bar{\lambda}_{\dot{B}}\Big) \d\bar{\Theta}^{\dot{B}}
{\bar{\l}}_{\dot{B}} 
\right],
$$
where the basis in the space of variations of $x$ and $y$ is 
chosen in the form   
\begin{equation}\plabel{var}
\begin{array}{l}
\displaystyle
i_\d \Pi^{A\dot{B}} \equiv  
 \d{x}^{A\dot{B}}
+ i \left( \d\Theta^{A} \bar{\Theta}^{\dot{B}}
- \Theta^{A} \d\bar{\Theta}^{\dot{B}}\right) \, ,
\\ \nonumber
\displaystyle
i_\d \Pi^{AB} \equiv 
\d{y}^{AB} - ~i~ \Theta^{(A}~\d {\Theta}^{B)}\, ,
\\  
\displaystyle
i_\d \bar{\Pi}^{\dot{A}\dot{B}}
\equiv \d {\bar{y}}^{\dot{A}\dot{B}}
- ~i ~\bar{\Theta}^{(\dot{A}} ~d{\bar{\Theta}}\,{}^{\dot{B})}\, .
\end{array}
\end{equation}

\subsubsection{$U(1)$ gauge symmetry of $a=0$ model}
Let us consider the variation of the action when the variations of all
fields
except for $\lambda_A$, $\bar\lambda_{\dot A}$ are zero
\begin{equation}\plabel{lv}
\delta S = \int 
\left[
\d\lambda_{A}
\Big(\bar{\lambda}_{\dot{B}} \Pi^{A\dot{B}} +
2a \lambda_{B}  \Pi^{AB}\Big)
+ \d \bar{\lambda}_{\dot{A}} \Big(\Pi^{B\dot{A}}\lambda_{B} +
 2{a}  \bar{\lambda}_{\dot{B}} \Pi^{\dot{A}\dot{B}}
\Big) \right].
\end{equation}
If the variation of $\lambda$ corresponds to local infinitesimal $U(1)$
rotations 
\begin{equation}\plabel{eqdllss}
\l^\prime_A (\tau )= \l_A e^{i\a (\tau )} , \qquad 
\bar{\l}^\prime_{\dot{A}}=\bar{\l}_{\dot{A}} e^{-i\a (\tau )}
\end{equation}
the Eq. \p{lv} takes the form
$$
\delta S = \int 
\left[ 
i\a(\tau)\lambda_{A}
\Big(\bar{\lambda}_{\dot{B}} \Pi^{A\dot{B}} +
2a \lambda_{B}  \Pi^{AB}\Big)
-i\a(\tau) \bar{\lambda}_{\dot{A}} \Big(\Pi^{B\dot{A}}\lambda_{B} +
 2{a}  \bar{\lambda}_{\dot{B}} \Pi^{\dot{A}\dot{B}}
\Big) \right].
$$
Such a variation vanishes at $a=0$. Hence at this value of $a$
 the $U(1)$ 
transformations \p{eqdllss} describe local symmetry of the model which
is inherent to the Ferber-Shirafuji formulation \cite{F78,S83} of 
the massless superparticle.

Note that for all values of the parameter $a$ the spinors $\lambda$ are
constants on the mass shell. Indeed, the equations of motion 
$$
{\d S \over \d {x}^{A\dot{B}}}=0, \qquad 
 {\d S \over \d {y}^{AB}}=0, \qquad 
{\d S \over \d  {\bar{y}}^{\dot{A}\dot{B}}}=0
$$
which follow from \p{dac(a)} have the form 
\begin{equation}\plabel{eqdll}
d(\l_A\bar{\l}_{\dot{B}})=0 , \qquad 
a~d(\l_A \l_B)=0 , \qquad
a~d(\bar{\l}_{\dot{A}} \bar{\l}_{\dot{B}})=0.
\end{equation}
In the framework of any twistor or twistor-like approach
\cite{Penrose} 
one assumes that the bosonic spinors parametrize  a projective space.
This
requirement
does not allow $\l$ to have all its components equal to zero
simultaneously.
 
Then in the generic case $a\not=0$ eqs. \p{eqdll} imply 
\begin{equation}\plabel{eqdll1}
d(\l_A)=0 , \qquad 
 d(\bar{\l}_{\dot{A}})=0,
\end{equation}
i.e. the bosonic spinor is constant on the mass shell 
\begin{equation}\plabel{eqdlls}
\l_A (\tau )= \l_A^0=const , \qquad 
\bar{\l}_{\dot{A}}=\bar{\l}_{\dot{A}}^0=const. 
\end{equation}

When $a=0$ only one equation is left in \p{eqdll} 
\begin{equation}\plabel{eqdll0}
a=0: \qquad d(\l_A\bar{\l}_{\dot{B}})=0 . \qquad 
\end{equation}
The general solution of \p{eqdll0} is 
 \begin{equation}\plabel{eqdlls0}
\l_A (\tau )= \l^0_A e^{i\tilde\a (\tau )} , \qquad 
\bar{\l}_{\dot{A}}=\bar{\l}^0_{\dot{A}} e^{-i\tilde\a (\tau )}.
\end{equation}
The arbitrary function  $\tilde\a (\tau )$ (whose presence in
\p{eqdlls0}
reflects the $U(1)$ gauge symmetry of the $a=1$ model)
can be gauged away by the
local $U(1)$ transformation \p{eqdllss}, and we are
again left with constant $\l$ on the mass shell.

\bigskip

\subsubsection{Fermionic variations and $\kappa$--symmetry}

Let us consider now  
the formula \p{dac(a)} with the
 variations of fermionic coordinates accompanied
by the following variations of $x$ and $y$
\begin{eqnarray}\plabel{varXZ}
& \d{x}^{A\dot{B}}= 
- i \left( \d\Theta^{A} \bar{\Theta}^{\dot{B}}
- \Theta^{A} \d\bar{\Theta}^{\dot{B}}\right) 
~\Rightarrow 
  & i_\d \Pi^{A\dot{B}}=0 ,  
\nonumber \\
& \d{y}^{AB} = ~i~ \Theta^{(A}~\d {\Theta}^{B)} ~~~~~~  \Rightarrow 
 & i_\d \Pi^{AB} =0 , 
\\  
& \d {\bar{y}}^{\dot{A}\dot{B}} 
= ~i ~\bar{\Theta}^{(\dot{A}} ~d{\bar{\Theta}}^{\dot{B})} 
~~~~~~  
\Rightarrow 
& i_\d \bar{\Pi}^{\dot{A}\dot{B}} =  0 . \nonumber
\end{eqnarray} 
The bosonic spinor $\lambda_\a$  remains unchanged. In such a
case eq. \p{dac(a)} takes the form
\begin{equation}\plabel{dfac(a)}
\delta S = \int 
\left[2i d{\Theta}^{A} \l_A 
\Big( \d\bar{\Theta}^{\dot{B}}
 \bar{\lambda}_{\dot{B}} + 
a \d{\Theta}^{{B}} \l_B \Big) 
+ 2i d{\bar{\Theta}}^{\dot{B}}{\bar{\l}}_{\dot{B}}
\Big(\d{\Theta}^{{B}} \l_B + {a} \d\bar{\Theta}^{\dot{B}}
 \bar{\lambda}_{\dot{B}} \Big) 
\right].
\end{equation}

We see that for $a\not=1$ only two out
 of four variations of the fermionic coordinates 
$\d{\Theta}^{{A}},  \d\bar{\Theta}^{\dot{A}} $ are effectively involved 
into the variation \p{dfac(a)} of the action \p{ac(a)}. 
This reflects the presence 
of local fermionic $\kappa$-- symmetry \cite{AL}
with {\sl two} independent parameters $\kappa=(\kappa_1 + i\k_2)$, 
$\bar{\kappa}=(\kappa_1 - i\k_2)$. The $\kappa$--transformations of the
coordinates are given by 
eq. \p{varXZ} and 
\begin{equation}\plabel{kappag}
\d{\Theta}^{A} = \kappa \l^A = (\kappa_1 + i\k_2) \l^A, \qquad 
\d\bar{\Theta}^{\dot{B}}
 = \bar{\kappa}\bar{\lambda}^{\dot{B}} = 
(\kappa_1 - i\k_2) \bar{\lambda}^{\dot{B}}. 
\end{equation}

At the critical point $a=1$ 
the number of independent $\kappa$--symmetries increases from two to 
{\sl three}, since in this case 
only one linear combination 
($\d{\Theta}^{{B}} \l_B + \d\bar{\Theta}^{\dot{B}}
\bar{\lambda}_{\dot{B}}$) 
of four real fermionic variations enters into
  the variation of the action 
\begin{equation}\plabel{dfac(1)}
a= 1~:~ \qquad  
\delta S = \int 
2i 
\Big(d{\Theta}^{{A}} \l_A +  d\bar{\Theta}^{\dot{A}}
 \bar{\lambda}_{\dot{A}}\Big) ~
\Big(  \d{\Theta}^{{B}} \l_B  
+ d\bar{\Theta}^{\dot{B}}
 \bar{\lambda}_{\dot{B}} 
\Big).
\end{equation}
Thus remaining three fermionic variations correspond to the local
fermionic  
symmetries of the $a=1$ model \cite{BL}. 

In order to present an explicit form of these {\sl three $\kappa$ 
symmetries} 
we should introduce an additional bosonic spinor  
$u_A$ such that    
\begin{equation}\plabel{hatu}
\l^A {u}_A = \bar{\l}^{\dot{A}}{\bar{u}}_{\dot{A}} =1. 
\end{equation}
Then one can perform the decomposition of the unit matrix in the 
 spinor space \footnote{The pair of Weyl spinors 
  $\lambda^{A}, u_{A}$  is analogous to the Newman--Penrose dyad 
\cite{NewmanPenrose63} widely used in General Relativity.}
\begin{equation}\plabel{ddec}
\d_B^{~A} = \l^A{u}_B -{u}^A \l_B, 
\qquad  \d_{\dot{B}}^{\dot{A}}=   
\bar{\l}^{\dot{A}} {\bar{u}}_{\dot{B}} - 
{\bar{u}}^{\dot{A}} \bar{\l}_{\dot{B}} 
\end{equation}
and use it to decompose the fermionic variation of $\Theta$.
 
As a result we find that the $\kappa$--symmetry transformations of the 
$a=1$ model are given by eqs. \p{varXZ} and 
\begin{equation}\plabel{kappa1}
\d{\Theta}^{A} = (\kappa_1 + i\k_2) \l^A + i \k_3{u}^A, 
 \qquad 
\d\bar{\Theta}^{\dot{B}}
 = 
(\kappa_1 - i\k_2) \bar{\lambda}^{\dot{B}} - 
i \k_3{\bar{u}}^{\dot{A}}. 
\end{equation}

\subsection{Hamiltonian analysis}

We  now turn to the Hamiltonian 
analysis of the model with the purpose of getting all the constraints
on the dynamics of the system, classifying them {\it a la} Dirac and
thus 
identifying all local symmetries of the model. 

For the case $a=1$ the analysis has been performed in \cite{BL}.
The generic model ($a\not =0$) has the same total number of constraints 
as the $a=1$ model, the only difference being that when the parameter
$a$
takes the value $a=1$ one of
the fermionic second--class constraints 
becomes the first--class constraint generating the third
$\kappa$--symmetry. 
So what we should  do is just to adapt the results
of the Hamiltonian analysis of \cite{BL} to the generic case.

The constraints corresponding to the 
case $a=0$ of the Ferber--Shirafuji superparticle are obtained from the
generic set of constraints by putting the canonical momenta for the
central charge coordinates identically equal to zero.

The canonical momenta of the generic system are
\begin{equation}\label{momenta}
{\cal P}_{{\cal M}} = { \partial L \over \partial \dot{q}^{{\cal M}} }
= (P_{A\dot{A}}, Z_{AB},
\bar{Z}_{\dot{A}\dot{B}};  P^A, \bar{P}^{\dot{A}};
\pi_{A}, \bar{\pi}_{\dot{A}}) ,
\end{equation}
\begin{equation}\label{brackets}
[ {\cal P}_{{\cal M}}, {q}^{{\cal N}} \}_P
= - (-1)^{{\cal M}{\cal N}}
[{q}^{{\cal N}}, {\cal P}_{{\cal M}} \}_P=  
\d _{{\cal M}}^{{\cal N}} ~:
\end{equation}
$$
[P_{A\dot{A}}, x^{B\dot{B}}]_P = \d_{A}^{B}\d_{\dot{A}}^{\dot{B}},
\qquad 
[Z_{AB}, y^{CD}]_P = 2 \d_{[A}^{C}\d_{B]}^{D}, \qquad 
[\bar{Z}_{\dot{A}\dot{B}}, \bar{y}^{\dot{C}\dot{D}}]_P = 
2 \d_{[\dot{A}}^{\dot{C}} \d_{\dot{B}]}^{\dot{D}}, \qquad 
$$  
$$
[P^{A}, \l_{B}]_P = \d_{B}^{A}, \qquad 
[\bar{P}^{\dot{A}}, \bar{\l}_{\dot{B}}]_P = \d^{\dot{A}}_{\dot{B}},
\qquad
$$
$$
\{\pi_{A}, \Theta^B \}_P= \d_{A}^{B}, \qquad 
\{ \bar{\pi}_{\dot{A}}, \bar{\Theta}^{\dot{B}} \}_P= 
\d_{\dot{A}}^{\dot{B}}. \qquad
$$

They satisfy the following set of constraints 
\begin{equation}\label{Phi1}
\Phi_{A\dot{B}} \equiv
P_{A\dot{B}} - \lambda_{A} \bar{\lambda}_{\dot{B}} = 0 ,
\end{equation}
\begin{equation}\label{Phi2}
\Phi_{A{B}} \equiv
Z_{A{B}} - a \lambda_{A} {\lambda}_{{B}} = 0 ,
\end{equation}
\begin{equation}\label{Phi3}
\bar{\Phi}_{\dot{A}\dot{B}} \equiv
\bar{Z}_{\dot{A}\dot{B}} - a 
\bar{\lambda}_{\dot{A}} \bar{\lambda}_{\dot{B}} = 0 ,
\end{equation}
\begin{equation}\label{PA=0}
P_{A}=0, \qquad
\bar{P}_{\dot{A}}  = 0 ,
\end{equation}
   
\begin{equation}\label{DA=0}
D_{A}\equiv - \pi_A + i P_{A\dot{B}} \bar{\Theta}^{\dot{B}} + i
Z_{AB} \Theta^B=0, \qquad
\end{equation}
\begin{equation}\label{bDA=0}
\bar{D}_{\dot{A}}\equiv  \bar{\pi} _{\dot{A}} - i \Theta^B  P_{B\dot{A}}
- i \bar{Z}_{\dot{A}\dot{B}} \bar{\Theta}^{\dot{B}} = 0.
\end{equation}

To separate the constraints \p{Phi1}--\p{bDA=0} 
into the first and second class
let us project them on  the bosonic spinors $\lambda$ and $u$ 
(Eqs. \p{hatu}, \p{ddec}). We get

\begin{equation}\label{B1}
 B_1 =
 {\l}^{A} {\bar{\l}}^{\dot{B}}
 P_{A\dot{B}}   = 0,
\end{equation}
\begin{equation}\label{B2}
 B_2 =
 {\l}^{A} {\bar u}^{\dot{B}}
 P_{A\dot{B}}  - \l^A{u}^B Z_{AB} = 0,
\end{equation}
\begin{equation}\label{B3}
 B_3 \equiv (B_2)^*=
 {u}^{A} {\bar{\l}}^{\dot{B}}
 P_{A\dot{B}}  - \bar{\l}^{\dot{A}} {\bar{u}}^{\dot{B}}
 \bar{Z}_{\dot{A}\dot{B}} = 0,
\end{equation}
\begin{equation}\label{B4}
 B_4 =
2 {u}^{A} \hat{{u}}^{\dot{B}}
 P_{A\dot{B}}  - {1\over a}{u}^A {u}^B Z_{AB}
 - {1\over a}{\bar{u}}^{\dot{A}} {\bar{u}}^{\dot{B}} 
\bar{Z}_{\dot{A}\dot{B}}
  = 0,
\end{equation}
\begin{equation}\label{B5}
 B_5 =
 {\l}^{A} {\bar{\l}}^{{B}}
 Z_{AB} = 0,
\end{equation}
\begin{equation}\label{B6}
 B_6 \equiv (B_5)^*=
 \bar{\l}^{\dot{A}} {\bar{\l}}^{\dot{B}}
 \bar{Z}_{\dot{A}\dot{B}} = 0,
\end{equation}
\begin{equation}\label{SB12}
 B_7\equiv {\l}^{A}{\bar{u}}^{\dot{B}}
 P_{A\dot{B}}  + \l^A {u}^B Z_{AB} = 0,  \qquad
 B_8\equiv {u}^{A} {\bar{\l}}^{\dot{B}}
 P_{A\dot{B}}  + \bar{\l}^{\dot{A}}{\bar{u}}^{\dot{B}}
 \bar{Z}_{\dot{A}\dot{B}} = 0,
\end{equation}
\begin{equation}\label{SB34}
 B_9\equiv {u}^A {u}^B Z_{AB} - a = 0, \qquad
B_{10}\equiv {\bar{u}}^{\dot{A}} {\bar{u}}^{\dot{B}}
\bar{Z}_{\dot{A}\dot{B}} -a
  = 0,
\end{equation}
\begin{equation}\label{SB56}
B_{11}\equiv i(\lambda^AP_{A}-\bar\l^{\dot A}\bar{P}_{\dot{A}}) =0,
\end{equation}
\begin{equation}\label{SB57}
B_{12}\equiv \lambda^AP_{A}+\bar\l^{\dot A}\bar{P}_{\dot{A}}=0,  
\end{equation}
$$
B_{13}\equiv u^AP_{A}=0,\qquad 
B_{14}\equiv \bar{u}^{\dot A}\bar{P}_{\dot{A}} = 0.
$$
\begin{equation}\label{F1}
 F_1 =
 {\l}^{A}
 D_{A} = 0,
\end{equation}
\begin{equation}\label{F2}
 F_2 \equiv (F_1)^*=
 \bar{\l}^{\dot{A}} {\bar{D}}_{\dot{A}}= 0,
\end{equation}
\begin{equation}\label{F3}
 F_3 =
 {u}^{A}
 D_{A} +
 {\bar{u}}^{\dot{A}} {\bar{D}}_{\dot{A}}= 0,
\end{equation}
\begin{equation}\label{SF1}
 F_4\equiv {u}^{A}
 D_{A} -
 {\bar{u}}^{\dot{A}} {\bar{D}}_{\dot{A}}= 0.
\end{equation}

For  arbitrary $a\not =0,1$ it can be checked that
the bosonic constraints \p{B1}--\p{B6} and the fermionic constraints
\p{F1} and \p{F2} belong to the first class, i.e. their Poisson brackets
with all constraints vanish on the constraint surface, and the
constraints
\p{SB12}--\p{SB57}, \p{F3} and \p{SF1} are second--class. When computing
the Poisson brackets of the constraints one should take into account
that,
because of the normalization condition \p{hatu}, the spinor $u^A$ 
 should be regarded as a variable depending on $\l_A$. The simplest way 
of taking this into account is 
to assume 
the
following Poisson (actually Dirac) brackets of $u^A$ 
with the $\lambda$--momentum $P_A$
\footnote{These brackets appear as Dirac brackets with respect to the 
pair of the second class constraints \p{hatu} and $u^AP^{(u)}_A = 0$, 
(and their complex conjugate pair), when the bosonic spinor $u$ is 
considered as 
an independent variable whose momentum is  constrained to be zero 
$P_A^{(u)} = 0$. Then it is not hard to verify that the new phase 
space variables $u^A, P^{(u)}_A$ do not introduce new redundant 
degrees of freedom into the system under consideration. }
$$
[P_A,u^B]_P=- u^Bu_A.
$$

Thus in the case $a\not = 0,1$ among 14 bosonic and 4 fermionic
constraints
6 bosonic and 2 fermionic constraints are of the first class and
8 bosonic and 2 fermionic constraints are of the second class.

The first class constraints generate 
local symmetries of the dynamical system.
For instance, the constraints \p{B1}, and $(B_5+B_6)$ of \p{B5}, \p{B6}
generate worldline reparametrizations of the coordinates $x$ and $y$.
The fermionic constraints \p{F1} and \p{F2} generate the
$\kappa$--symmetry
transformations \p{varXZ} and \p{kappag}.

Each  first class constraint reduces the number of independent phase
space
variables by  two, while each second class constraint
eliminates only one degree of freedom. Hence, in the case $a\not =0,1$
the phase space of $2\times (4+6+4)=28$ bosonic and $2\times 4=8$
fermionic
canonical variables of the system is reduced to
\begin{equation}\label{H(a)}
a\not= 0, 1: \qquad n_{ph}=\hbox{{\bf 8$_b$ + 2$_f$}}
\end{equation}
i.e. we get 
 eight bosonic and two fermionic physical degrees of freedom.

In order to see how at $a=1$ the fermionic second--class constraint
\p{F3}
transforms
into the first--class constraint generating the third $\kappa$--symmetry
\p{kappa1} let us  consider the Poisson bracket of the constraint \p{F3}
with
itself
\begin{equation}\label{F3PB}
\{F_3,F_3\}_P=2(a-1).
\end{equation}
When $a\not =1$ the r.h.s. of \p{F3PB} is nonzero and hence this
constraint is
second class, but at $a=1$ the r.h.s. of \p{F3PB} vanishes. 
Since $F_3$ weakly 
commutes with all other constraints, at this critical value of
$a$ 
we obtain one more fermionic first class constraint, and 
 we achieve
the reduction of the number of
independent
fermionic physical degrees of freedom from two to one
\begin{equation}\label{H(a1)}
a=1: \qquad n_{ph}=\hbox{{\bf $8_b$ + $1_f$}}.
\end{equation}

Finally, when $a=0$ the tensorial coordinates $y$ disappear from the
action \p{ac(a)}, and in Eqs. \p{B1}--\p{SF1} we must put to zero
their canonical momenta $Z$. The remaining set of the constraints takes
the following form
\begin{equation}\label{B10}
 B_1 =
 {\l}^{A} {\bar{\l}}^{\dot{B}}
 P_{A\dot{B}}   = 0,
\end{equation}
\begin{equation}\label{B20}
 B_2 =
 {\l}^{A} {\bar u}^{\dot{B}}
 P_{A\dot{B}} = 0,
\end{equation}
\begin{equation}\label{B30}
 B_3 \equiv (B_2)^*=
 {u}^{A} {\bar{\l}}^{\dot{B}}
 P_{A\dot{B}}  = 0,
\end{equation} 
\begin{equation}\label{B40}
 B_4 =
{u}^{A} \hat{{u}}^{\dot{B}}
 P_{A\dot{B}}  - 2 = 0,
\end{equation}
\begin{equation}\label{SB560}
B_{5}\equiv i(\lambda^AP_{A}-\bar\l^{\dot A}\bar{P}_{\dot{A}}) =0,
\end{equation}
\begin{equation}\label{SB570}
B_{6}\equiv \lambda^AP_{A}+\bar\l^{\dot A}\bar{P}_{\dot{A}}=0,  
\end{equation}
$$
B_{7}\equiv u^AP_{A}=0,\qquad 
B_{8}\equiv \bar{u}^{\dot A}\bar{P}_{\dot{A}} = 0.
$$
\begin{equation}\label{F10}
 F_1 =
 {\l}^{A}
 D_{A} = 0,
\end{equation}
\begin{equation}\label{F20}
 F_2 \equiv (F_1)^*=
 \bar{\l}^{\dot{A}} {\bar{D}}_{\dot{A}}= 0,
\end{equation}
\begin{equation}\label{F30}
 F_3 =
 {u}^{A}
 D_{A} +
 {\bar{u}}^{\dot{A}} {\bar{D}}_{\dot{A}}= 0,
\end{equation}
\begin{equation}\label{SF10}
 F_4\equiv {u}^{A}
 D_{A} -
 {\bar{u}}^{\dot{A}} {\bar{D}}_{\dot{A}}= 0.
\end{equation}
These are the constraints of the Ferber--Shirafuji formulation of the
superparticle which has been analyzed in detail in a number of papers
\cite{S83,14,15,ES,17}.
Now two bosonic constraints \p{B10} and \p{SB560} are first--class and
other six are second--class, while, as in the generic case  $a\not =1$,
two of the fermionic constraints are first--class \p{F10}, \p{F20} and
two are second--class \p{F30}, \p{SF10}.

Therefore, the number of independent phase--space physical degrees 
of freedom of
the standard $N=1$, $D=4$ superparticle consists of six bosonic and two
fermionic variables
\begin{equation}\label{H(a0)}
a= 0: \qquad n_{ph}=\hbox{{\bf $6_b$ + $2_f$}}.
\end{equation}

In the next section we shall show that the independent phase--space
physical
degrees of freedom \p{H(a)}, \p{H(a1)} and \p{H(a0)}
of the generic superparticle model can be covariantly described by
$OSp(2|8)$ (or $OSp(1,1|8)$), $OSp(1|8)$ and $SU(2,2|1)$ supertwistors,
respectively.

\section{Supertwistor transform. 
$OSp(2|8)$, $OSp(1|8)$ and $SU(2,2|1)$ supertwistors}
\setcounter{equation}{0}

Let us integrate the action \p{ac(a)} by parts and neglect the boundary
term.
 The result is
\begin{equation}\plabel{ac(a)tw}
S = - \int \left( \mu^A d\l_A +
\bar{\mu}^{\dot{A}}d\bar{\l}_{\dot{A}}\right)
- i\int \left( \chi d \bar{\chi}+ \bar{\chi}d\chi +a \chi d\chi + 
{a} \bar{\chi}d\bar{\chi}
\right)
\end{equation}
or 
$$
S = - \int \left( \mu^A d\l_A +
\bar{\mu}^{\dot{A}}d\bar{\l}_{\dot{A}}\right)
- 2i\int \left( (1+a)\chi_1 d{\chi}_1+ (1-a){\chi}_2d\chi_2 
\right)
$$
\begin{equation}\plabel{ac(a)tw2}
 = - \int \left( \mu^A d\l_A +
\bar{\mu}^{\dot{A}}d\bar{\l}_{\dot{A}}\right)
- 2i\int \bar\chi(a)d{\chi}(a),
\end{equation}
where 
\begin{eqnarray}\plabel{Pcorr(a)b}
\mu^A= x^{A\dot{B}}\bar{\l}_{\dot{B}} + 2a y^{AB}\l_B 
+ i \Theta^A [(\bar{\Theta}\bar{\l}) + a ({\Theta}{\l})], 
\qquad \\ \nonumber 
\bar{\mu}^{\dot{A}} = \l_B x^{B\dot{A}} + 2\bar{a}
\bar{y}^{\dot{A}\dot{B}}
\bar{\l}_{\dot{B}} + i \bar{\Theta}^{\dot{A}} 
[({\Theta}{\l})+ \bar{a} (\bar{\Theta}\bar{\l})],  
\end{eqnarray}
\begin{equation}\plabel{Pcorr(a)f}
\chi = ({\Theta}{\l})\equiv {\Theta}^A {\l}_A, \qquad 
\bar{\chi}= (\bar{\Theta}\bar{\l})\equiv
 \bar{\Theta}^{\dot{A}} \bar{\l}_{\dot{A}}, 
\end{equation}
\begin{equation}\plabel{chi}
\chi_1 = {1 \over 2} 
( \chi + \bar{\chi}),\qquad \chi_2 = {i \over 2} ( \bar{\chi}- \chi).
\end{equation}
\begin{equation}\plabel{chia}
\chi(a)=\sqrt{(1+a)}\chi_1+i\sqrt{(1-a)}\chi_2,\quad
\bar\chi(a)=\sqrt{(1+a)}\chi_1-i\sqrt{(1-a)}\chi_2.
\end{equation}
(Note that $\chi(a)$ and $\bar\chi(a)$ are complex conjugate to each
other
only for $a<1$, while for $a\ge 1$ they are real spinors).

Thus, in the generic case $a\not=0,1$ one can reformulate
  the dynamical system 
  in terms of 
8 bosonic variables $\l_A, \mu^A ; \bar{\l}_{\dot{B}},
\bar{\mu}^{\dot{B}}$
and two real fermionic variables 
$\chi_1,~ \chi_2$. 
These variables can be regarded as components of a real 
$(8,2)$ component supertwistor (cf. with \cite{BL})
\begin{equation}\plabel{stw(a)}
{Y}_{{\cal A}} = (y_1, \ldots , y_8; \chi_1,\chi_2 ) 
= (\l^\a, \mu^\a, \chi_1,\chi_2)
\end{equation}
where $\l^\a$ and $\mu^\a$ are real Majorana 
spinors formed of the  Weyl spinors 
$$
\l^\a ~\Leftrightarrow ~ \left(\matrix{\l_A \cr
\bar{\l}^{\dot{A}}}\right), 
\qquad \mu^\a ~\Leftrightarrow ~ \left(\matrix{\mu_A \cr
\bar{\mu}^{\dot{A}}}
\right)
$$

One can  write the action \p{ac(a)tw2} in the form 
 \begin{equation}
 \plabel{actw(a)2}
S = - {1 \over 2} \int d \tau {Y}_{{\cal A}} G^{{\cal A} {\cal B}}
\dot{Y}_{{\cal B}}
\end{equation}  
where 
\begin{equation}
 \plabel{OSpm(a)}
  G^{{\cal A}{\cal B}} = \left( \matrix{
   \om^{(8)} & 0 \cr
   0 & i \om^{(2)}
 \cr} \right) =
   \left(
  \matrix{
  \left({\matrix{
   0_2 & I_2 & 0_2 & 0_2  \cr
  -I_2 & 0_2 & 0_2 & 0_2   \cr
   0_2 & 0_2 & 0_2 & I_2   \cr
  0_2 & 0_2 & {-I}_2 & 0_2 \cr  }}\right)
    &  0   \cr
   0 &  i
\left(
  \matrix{ 2(1+a) & 0 \cr 
           0  & 2(1-a) }\right)
\cr}           
\right)\, .
\end{equation}
$\om^{(8)}$ is the $Sp(8)$ invariant simplectic metric.

When $a\not = 1$  we can rescale the fermionic variables ${\chi}_1$ and 
${\chi}_2$, i.e. 
 multiply them, respectively, by  $\sqrt{1+a}$ and $\sqrt{|1-a|}$.
This results in the following form of the metric $\om^{(2)}$ in
\p{OSpm(a)}
\begin{equation}
 \plabel{mfer(a)+}
\om^{(2)} 
\rightarrow ~~~ \om^{(2)}= 
   2 \left(
   \matrix{ 1 & 0 \cr 
              0   & 1 }\right)\qquad for~a<1 
\end{equation}
or 
\begin{equation}
 \label{mfer(a)-}
\om^{(2)} 
\rightarrow ~~~ \om^{(2)} = 
   2 \left(
   \matrix{ 1 & 0 \cr 
              0   & -1 }\right)\qquad for~a>1. 
\end{equation}
We see that the symmetry group of
the fermionic sector of the metric 
\p{OSpm(a)} is $SO(2)=U(1)$ when $a<1$ and 
$SO(1,1)$ when $a>1$. 

Hence, when $a<1$ the complete symmetry group of the 
metric \p{OSpm(a)} is the supergroup $OSp(2|8)$, while in the case $a>1$
the symmetry group becomes $OSp(1,1|8)$.  The supertwistors \p{stw(a)}
transform under the fundamental representations of these supergroups.

When $a=1$  the 
metric becomes degenerate 
\begin{equation}
 \plabel{mfer(a)1}
\om^{(2)} 
\rightarrow ~~~ \om^{(2)}_{a=1} = 
   2 \left(
   \matrix{ 1 & 0 \cr 
              0   & 0 }\right).
\end{equation}
This reflects the absence of the
second fermionic variable $\chi_2$ from the action \p{ac(a)tw2}.
Thus at $a=1$ the supergroups $OSp(2|8)$ and $OSp(1,1|8)$ reduce to
$OSp(1|8)$ with the corresponding supertwistor representation having
one real fermionic component (cf. with \cite{BL})

\begin{equation}\plabel{stw(1)}
{Y}_{{\cal A}} = (y_1, \ldots , y_8; \chi_1).
\end{equation}

Consider now the case $a=0$. At $a=0$ the action has the same form as
for $a<1$ and hence is formally $OSp(2|8)$ invariant. But, as we have
seen in the previous section, at this critical point the model acquires
additional local $U(1)$ symmetry, which must have its counterpart in the
supertwistor description, i.e. there should appear a first--class
constraint
on the supertwistor variables which generates this symmetry. In
 order to identify 
this constraint 
we use the defining relations \p{Pcorr(a)b}
and
  consider the following
  bilinear 
   combination of supertwistor
components
 for an arbitrary value of $a$ 

\begin{equation}\plabel{U(1)(a)}
\mu^A(a) \l_A - \bar{\mu}^{\dot{A}}(a) \bar{\l}_{\dot{A}} + 2i 
\bar{\chi} \chi= 
2a \l y \l -  2\bar{a} \bar{\l} \bar{y}\bar{\l}.
\end{equation}
At $a=0$ eq. \p{U(1)(a)} does 
not involve central charge coordinates $y$ and thus 
 we obtain the pure supertwistor constraint 
\begin{equation}\plabel{U(1)(0)}
\mu^A(0) \l_A - \bar{\mu}^{\dot{A}}(0) \bar{\l}_{\dot{A}} + 2i 
\bar{\chi} \chi= 0.
\end{equation}
Hence, at $a=0$ for the action \p{ac(a)tw} to be equivalent to 
 \p{ac(a)} 
   it  must be supplemented with 
the (first--class) constraint \p{U(1)(0)} introduced through a
Lagrange multiplier term
\begin{equation}\plabel{ac(0)tw}
a=0: \qquad 
S = - \int \left( \mu^A d\l_A +
\bar{\mu}^{\dot{A}}d\bar{\l}_{\dot{A}}\right)
- i\int \left( \chi d \bar{\chi}+ \bar{\chi}d\chi  \right) + i 
\int d\tau \L 
(\mu \l - \bar{\mu} \bar{\l} + 2i 
\bar{\chi} \chi) 
\end{equation} 

The constraint \p{U(1)(0)} generates the $U(1)$ gauge symmetry 
appearing only in
 the $a=0$ case. This constraint introduces the complex structure 
and thus breaks the $OSp(2|8)$ symmetry of the $a<1$ model 
down to $SU(2,2|1)$. As a result one gets 
the Ferber-Shirafuji formulation \cite{F78,S83} of a
conventional massless 
superparticle \cite{casal,vt,BS} in terms of  $SU(2,2|1)$
supertwistors 
$$
{\cal Z}_{{\cal A}} = (\l_A, \bar{\mu}^{\dot{A}}, \chi ), \qquad 
\bar{{\cal Z}}^{{\cal A}} = ( \mu^A, \bar{\l}_{\dot{A}},
\bar{\chi} ),
$$
\begin{equation}\plabel{ac(0)tw1}
a=0: \qquad 
S = - \int \left(  \bar{{\cal Z}}^{{\cal A}}d{\cal Z}_{{\cal A}}
\right)
+ i 
\int d\tau \L  (\bar{{\cal Z}}^{{\cal A}}{\cal Z}_{{\cal A}} - s),
\end{equation} 
where the constant $s$ has been introduced in order
  to have the possibility
of 
describing massless superparticles with nonzero (super)helicity 
\cite{Penrose} (see \cite{ES,B90} and references therein for details).

\subsection{Quantization of the supertwistor model}
\subsubsection{Canonical supertwistor quantization}
The quantization of the 
dynamical system \p{ac(a)tw2} with $a\not=0,1$ is quite straightforward. 
The action is of the first order form, therefore  $\mu, \bar{\mu}$ 
 should be identified with the canonical momenta conjugate to 
$\l_A, \bar{\l}_{\dot{A}},$ and $2i\bar{\chi}(a)$ is the
momentum
conjugate to $\chi(a)$ (remember that $\chi(a)$ and $\bar{\chi}(a)$ are
defined by \p{chia}). The canonical Poisson brackets are
\begin{equation}\plabel{twbrack}
[\mu_A, \l^B ]_P = \d_A^{~B}, \qquad 
[\bar{\mu}_{\dot{A}}, \bar{\l}^{\dot{B}}]_P = \d_{\dot{A}}^{\dot{B}},
\qquad \{\bar\chi(a),\chi(a)\}_P=-{i \over 2}.
\end{equation}  
At the quantum level the dynamical variables become operators, and the
Poisson brackets are replaced by (anti)commutators 
($[..., ...]_P~\rightarrow i[...,...]$, 
$\{ , \}_P ~\rightarrow -i\{ ,\}$).).
For instance,
in the `coordinate' representation the momenta are the derivatives of 
corresponding coordinates
\begin{equation}\plabel{twqu}
\mu^A = i {\partial \over \partial \l_A} , \qquad 
 \bar{\mu}^{\dot{A}} = i {\partial \over \partial \bar{\l}_{\dot{A}}}, 
\qquad 
\bar{\chi}(a) = - {1 \over 2} {\partial \over {\partial \chi(a)}}. 
\end{equation}

The canonical Hamiltonian of the system vanishes identically.

The wave function of the system in the supertwistor
`coordinate' representation is  
\begin{equation}\plabel{twvfa<1}
a\not =1: 
\Phi (\l_A, \l_{\dot{A}}, \chi(a) )= \phi (\l_A, \l_{\dot{A}}) +  
i \chi(a) \psi (\l_A, \l_{\dot{A}})
\end{equation}
and the spectrum of quantum states is described by 
one bosonic and one fermionic function depending on  
Weyl spinor variables. 

%

At $a=1$ $\chi(1)$ becomes a real Clifford variable and 
the field \p{twvfa<1} becomes  a Clifford algebra valued function.
We shall discuss this case in detail in Subsections 3.1.4, 5.2. and
Appendix.

To understand what kind of physical states are described by the
function \p{twvfa<1} in the case $a\not=0,1$ we shall first consider the 
well known case $a=0$.

\subsubsection{ $a=0$} 

At $a=0$ the dynamics of the system is subject to the 
first--class constraint \p{U(1)(0)} which at the quantum level is
imposed
on the wave function \p{twvfa<1} \footnote{In this section we basically
follow the quantization procedure of references \cite{15}. The
operator $D=D^{(0)} + \chi {\partial \over \partial \chi} $ is
the superhelicity operator.}
\begin{equation}\plabel{twD0} 
  (D^{(0)}  + \chi {\partial \over {\partial \chi} }- s) 
\Phi (\l_A, \bar\l_{\dot{A}}; \chi ) =0, 
\end{equation}
where 
$$
D^{(0)}= \l_A {\partial \over \partial \l_A} - 
\bar{\l}_{\dot{A}} {\partial \over \partial \bar{\l}_{\dot{A}}} 
$$
is the supertwistor representation of the bosonic part of the $U(1)$ 
generator, $\chi=\chi(0)$ (see Eqs. \p{Pcorr(a)f}--\p{chia}) and $s$ is
an
integer
constant which appears due to the ambiguity in ordering the operators in
\p{twD0}
(see \cite{Penrose,ES,B90} and refs. therein for details, 
here we only note that the quantization of $s$ follows from 
the requirement for the wave function to be 
single valued).  The (half)integer values of $s/2$ describe helicities
of massless quantum states.

Let us consider first  
the case $s=0$. 
 Eq. \p{twD0} requires the 
bosonic and fermionic components of the superfield \p{twvfa<1}
to be homogeneous functions of $\l, \bar{\l}$ of the degree $0$ and
$-1$, 
respectively,
\begin{equation}\plabel{twD0c} 
  D^{(0)} \phi (\l_A, \l_{\dot{A}}) =0, \qquad 
 D^{(0)} \psi (\l_A, \l_{\dot{A}}) = - \psi (\l_A, \l_{\dot{A}}) 
 \qquad 
\end{equation}
The solution is \footnote{
A rigorous approach \cite{B90,ZF} 
consists in the consideration of the decomposition 
of the wave function $\phi (\l_A , \bar{\l}_{\dot{A}})$ 
in the basis of the functions on {\bf C}$^2- \{ 0\}$ formed by 
homogeneous infinite-differentiable functions 
$ 
\phi_{\nu_1, \nu_2} (z\l_A , \bar{z}\bar{\l}_{\dot{A}})
= z^{\nu_1}\bar{z}^{\nu_2} \phi_{\nu_1, \nu_2} (\l_A ,
\bar{\l}_{\dot{A}})
$
of a homogeneity index $\chi = (\nu_1 , \nu_2 )$ \cite{Gelfand}. 
The homogeneous functions are defined by the Mellin transformation
$$
\phi_{\nu_1, \nu_2} (\l_A ,\bar{\l}_{\dot{A}})
= {i \over 2} \int dz d\bar{z}~z^{\nu_1+1}\bar{z}^{\nu_2+1}
\phi (z\l_A , \bar{z}\bar{\l}_{\dot{A}}).
$$
The decomposition 
$$
\phi (\l_A , \bar{\l}_{\dot{A}}) = 
\Sigma^{{}^{+\infty}}_{{}^{-\infty}} \int\limits^{+\infty}_{-\infty}
d\rho
\phi_{(n+i\rho )/2, (-n+i\rho  )/2} (\l_A , \bar{\l}_{\dot{A}})
$$
can be substituted 
into Eq. \p{twD0c} instead of the power series in $\l_A$, $\l_{\dot{A}}$ 
to obtain the general solution. 
We refer the reader to \cite{B90,ZF} 
for further details and to \cite{Gelfand} for an excellent presentation
of 
related mathematics, and, for simplicity, use a 
physical 'shortcut' of the rigorous approach.   
}

\begin{equation}\plabel{twD0s} 
  \phi= \phi_0 (p_m) , \qquad 
 \psi = \bar{\l}^{\dot{A}}{\bar\psi}_{\dot{A}} (p_m)
\end{equation}
where, by definition, $p_m$ is a light-like vector composed of $\l$ 
and $\bar\l$ (see also  \p{Phi1})        
\begin{equation}\plabel{twqCP} 
p_{A\dot{A}} = p_m \s^m_{A\dot{A}}=  \l_A \bar{\l}_{\dot{A}}. 
 \qquad 
\end{equation}

We see that the spectrum of the Ferber--Shirafuji model at $s=0$ 
consists of a massless $N=1$, $D=4$ (anti)chiral supermultiplet 
containing a complex
scalar field of zero helicity and a Weyl fermion field of helicity 
$-{1\over 2}$. This supermultiplet
can be described  either 
by the  set of bosonic and fermionic wave functions depending on the
bosonic Weyl spinor variables $\lambda_{A}, \lambda_{\dot{A}}$ 
 in accordance with the formula
\p{twD0s}-\p{twqCP}, or as a set of the unrestricted bosonic scalar
function 
$\phi_0 (p_m)$ and the fermionic spinor function
$\bar{\psi}_{\dot{A}}(p_m)$
depending on the light-like vector $p_mp^m=0$ which we identify
  with the momentum of the massless superparticle. 
In such a way we establish the relation of the supertwistor
formulation with the space--time description of the massless  superparticle, 
and this dual  
description can be extended to the case of more general
model with nonvanishing central charge coordinates. 

Finally, let us consider the case of the nonvanishing 
operator ordering constant $s$ in \p{twD0} which we shall call
the superhelicity parameter, characterizing the helicity
properties of the superfield solutions. 

The component form of the constraint \p{twD0} now reads  
\begin{equation}\plabel{twD0cs} 
  D^{(0)} \phi (\l_A, \l_{\dot{A}}) = s \phi (\l_A, \l_{\dot{A}}),
\qquad 
 D^{(0)} \psi (\l_A, \l_{\dot{A}}) = (s-1) \psi (\l_A, \l_{\dot{A}}) 
 \qquad 
\end{equation}

For integer $s>0$ the solution of \p{twD0cs}
is 
 \begin{equation}\plabel{twD0ss} 
  \phi= \l^{A_1} \ldots  \l^{A_s}\phi_{A_1 \ldots A_s} (p_m) , \qquad 
 \psi = \l^{A_1} \ldots  \l^{A_{s-1}} {\psi}_{A_1 \ldots A_{s-1}} (p_m)
\end{equation}
We thus obtain supermultiplets whose 
components have the helicities $s/2$ and $s/2-1/2$, respectively.
The choice of the statistics of the superfields \p{twD0ss} 
should be made in accordance with  the general spin--statistics theorem, 
such that for the even values of $s$ (integer superhelicities) the
superfields 
\p{twD0ss} are bosonic and 
for odd $s$ (half-integer superhelicities) they are fermionic.

Notice that the Grassmann parity of the superfield $\Phi$ \p{twD0}
(and its components $\phi$ and $\psi$) is related to the parity of
$\Phi(\lambda,\bar\l,\chi)$ under the change   $\lambda 
~\rightarrow~-\l$  ($\lambda$--parity)
  which implies $\chi~\rightarrow~-\chi$. 
   If
$\Phi(- \lambda,- \bar\l,- \chi)= \Phi(\lambda,\bar\l,\chi)$
 then
 from \p{twD0ss} follows that $s$ is even (integer superhelicities)
and 
  such a superfield 
  is Grassmann--even ($\phi$ is bosonic and $\psi$ is fermionic).
  Analogously if 
 the superfield 
$\Phi(\lambda,\bar\l,\chi)$ changes the sign under the
$\l$--parity, then $s$ is odd (half-integer superhelicities) and
the superfield
 is Grassmann--odd ($\phi$ is fermionic, and $\psi$ is bosonic).

For integer $s<0$ the solution of \p{twD0cs}
is 
 \begin{equation}\plabel{twD0ss1} 
  \phi= \l^{\dot{A}_1} \ldots  \l^{\dot{A}_{-s}}
\bar{\phi}_{\dot{A}_1\ldots  \dot{A}_{-s}}
\qquad 
 \psi = \l^{\dot{A}_1} \ldots  \l^{\dot{A}_{-s+1}}
\bar{\psi}_{\dot{A}_1\ldots  \dot{A}_{-s+1}}
\end{equation}
and thus the spectrum of the quantum states of the model  
 is represented by  a supermultiplet of helicity $(-s/2,(-s+1)/2)$.

\subsubsection{$a\not = 0,1$}

Let us return  to the generic $a\not= 0,1$ models. 
Their spectrum is defined by 
arbitrary scalar bosonic and fermionic functions of the 
Weyl bosonic spinors  
$\phi (\l_A, \l_{\dot{A}})$ and $\psi (\l_A, \l_{\dot{A}})$ which, 
in contrast
to the $a=0$ case, are not subject to any constraints. 
The bosonic spinor components can be regarded to be defined through 
the components of the light-like vector 
$p_mp^m=0$ \p{twqCP} up to the phase transformations 
\begin{equation}\plabel{phase}
\l_A ~\rightarrow~ e^{i\alpha (\tau)}\l_A, \qquad  
\l_{\dot{A}}~ \rightarrow~ e^{-i\alpha (\tau)}\l_{\dot{A}}.
\end{equation}
Thus for $a\not=0$ we can consider the bosonic and fermionic wave
functions 
to depend
on the light-like vector  and a $U(1)$ angle variable $\alpha \sim
\alpha + 2\pi k$ 
\begin{equation}\plabel{twa<1rep}
\phi (\l_A, \l_{\dot{A}})= \phi (p_m, \a ), \qquad 
\psi (\l_A, \l_{\dot{A}})= \psi (p_m, \a ). 
\end{equation}

Hence, in contrast to the Ferber--Shirafuji model, the wave function 
of the generic dynamical system  
\p{ac(a)tw} with $a\not=0$ depends on 
{\sl one additional variable} which  {\sl parametrizes a compact
manifold} 
$U(1)=S^1$. 
This means that the functions $\phi$ and $\psi$, as the single valued 
functions, can be expanded in the Fourier series
\begin{equation}\plabel{twa<1ser0}
\phi (p_m, \a )= \S_{k \in {\hbox{{\bf Z}}} }
e^{ik\a } \phi_k (p_m), \qquad 
\psi (p_m, \a )= \S_{k \in {\hbox{{\bf Z}}}} 
e^{ik\a } \psi_k (p_m).
\end{equation}  

The meaning of this series expansion becomes clear if we use  the 
Lorentz--covariant 
representation of $\phi$ and $\psi$ as single valued functions 
of  $\l_A$ \p{twvfa<1} where 
$\l_A\bar\l_{\dot A}$
are replaced by  
$p_m$. Then the series \p{twa<1ser0}
acquires the form 
\begin{equation}\plabel{sphi}
\phi (\l, \bar{\l} )= \phi^0 (p_m)+ \S_{k \in {\hbox{{\bf Z}}_+}} 
\left(\l^{A_1}\ldots \l^{A_k} \phi_{A_1\ldots A_k} (p_m) + 
\bar{\l}^{\dot{A}_1} \ldots \bar{\l}^{\dot{A}_k} 
\bar{\phi}_{\dot{A}_1 \ldots \dot{A}_k} (p_m) 
\right),  
\end{equation}  
$$
\psi (\l, \bar{\l} )= \psi^0 (p_m)+ \S_{k \in {\hbox{{\bf Z}}_+}} 
\left(\l^{A_1}\ldots \l^{A_k} \psi_{A_1\ldots A_k} (p_m) + 
\bar{\l}^{\dot{A}_1} \ldots \bar{\l}^{\dot{A}_k} 
\bar{\psi}_{\dot{A}_1 \ldots \dot{A}_k} (p_m) 
\right).
$$

We therefore conclude that the
 most general solution of the  model with $a\not = 0$ describes
an infinite  doubly degenerate spectrum
of massless fields of an arbitrary 
helicity, with  
the additional
compact $S^1$--coordinate 
 in the momentum space  conjugate to the discrete helicity
variable.

\bigskip 

If we assume the validity of spin--statistics theorem the
bosonic fields should have positive $\lambda$--parity, and fermionic
fields should have odd $\lambda$--parity. 
Thus, the $\l$-even part $
\Phi_+ (\l, \bar{\lambda}, \chi) 
\equiv  \Phi_+ (-\l, -\bar{\lambda}, -\chi) =
\phi_+ (\lambda, \bar{\lambda}) + i \chi \psi_-(\lambda, \bar{\lambda})
$
of the general superfield solution $
\Phi (\l, \bar{\lambda}, \chi) = 
\phi(\lambda, \bar{\lambda}) +  i \chi \psi (\lambda, \bar{\lambda})
$ (see 
\p{twvfa<1} and \p{sphi}) should be regarded as 
bosonic (i.e. Grassmann--even). Consequently this  implies that the
wave function $\phi_{+}(\lambda, \bar{\lambda})$ has positive
 $\lambda$--parity (even powers of $\lambda$) i.e. it  is
bosonic,  
and the fermionic wave
function $\psi_{-}(\lambda, \bar{\lambda})$ has negative
$\lambda$--parity (odd powers of $\lambda$).
Another sector of the full quantum state spectrum
  is described by the fermionic  
$\l$--odd superfield 
$
\Phi_- (\l, \bar{\lambda}, \chi) 
\equiv - \Phi_- (-\l, -\bar{\lambda}, -\chi) =
\phi_- (\lambda, \bar{\lambda}) + \chi \psi_+(\lambda, \bar{\lambda})
$ which is composed of the fermionic $\l$-odd field $\phi_- 
(\lambda, \bar{\lambda})$
and the bosonic $\l$--even field $\psi_+ (\lambda, \bar{\lambda})$.

We, therefore, see that in order to obtain physically meaningful
solutions described by the superfields $\Phi_{+}
(\lambda,\bar{\lambda}, \chi)$ and $\Phi_{-} (\lambda,
\bar{\lambda},\chi)$ with definite Grassmann parity one should
divide the general solution \p{twvfa<1} into two parts with even and odd
$\lambda$--parity. Note that the superfield
solutions with definite even/odd $\lambda$--parity have the even/odd
superhelicities, but each of them contains a complete
nondegenerate spectrum of states with both even and odd helicities.


It is instructive to compare the consequences of the presence of the 
 ``internal'' compact coordinate in our case and in Kaluza--Klein
theories.
In the Kaluza--Klein theories the compact variables arise in an
extension of space--time with extra directions and
lead to the quantization of corresponding ``internal''
  momenta in the extended momentum space.
The  ``internal'' 
quantized momenta describe masses and gauge charges of Kaluza--Klein
fields
in the dimensionally reduced theory 
\footnote{It is worth mentioning that `usual' Lorentz--scalar 
central charges 
can be interpreted as Kaluza-Klein momenta \cite{ALcch}}.
In our case 
the compactification is achieved by expressing the generalized momenta 
in terms of bosonic spinor (twistor) componenta. Thus, 
we have the opposite situation: 
the compact ``internal'' manifold is
in the extended 
   {\sl momentum} (twistor) space and a {\sl quantized (discrete)}
central charge coordinate is in the extended {\it coordinate} space
 (space--time + central charge coordinates).
The Fourier transform of the
compact ``internal" momentum 
results in the discrete values of the conjugate coordinate, which 
are described 
by an integer $s$. From the physical point of view the (half)integer number
$s/2$ 
describes the possible helicities of the massless quantum states.


The quantum states of our model form a reducible (infinite--dimensional)
representation of target space supersymmetry. 
Indeed, as the bosonic spinor is inert under global supersymmetry, 
the fields \p{sphi}
can be collected into the superfield series expansion with each
term having definite superhelicity  
\begin{equation}\plabel{sPhi}
\Phi (\l^{A}, \bar{\l}^{\dot{A}}, \chi(a) )= \Phi^0 (p_m,\chi (a) ) 
+ \S_{k \in {\hbox{{\bf Z}}_+}} 
\l^{A_1}\ldots \l^{A_k} \Phi_{A_1 \ldots A_k} (p_m,\chi(a)) +
\end{equation}  
$$
+ \S_{k \in {\hbox{{\bf Z}}_+}} 
\bar{\l}^{\dot{A}_1} \ldots \bar{\l}^{\dot{A}_k} 
\bar{\Phi}_{\dot{A}_1 \ldots \dot{A}_k} (p_m, \chi(a)) .
$$
It is easy to see that each term  is separately invariant under
supersymmetry 
\footnote{Remember that the supersymmetry transformations 
of the superfield \p{sPhi} are generated by the transformations of
$\chi$
as functions of $\Theta$ (eqs. \p{Pcorr(a)f}).}.

In the case $a=0$  the 
additional $U(1)$ constraint \p{twD0} appears. It 
singles out one irreducible superfield with a definite superhelicity
out of the infinite series \p{sPhi}.

\subsubsection{$a=1$}

Consider now a peculiarity of the model at $a=1$. In this case
the action \p{ac(a)tw},  \p{ac(a)tw2} contains only one real fermionic
variable $\chi_1$. The corresponding term in the action
is
\begin{equation}\plabel{chi1}
S_\chi=-4i\int \chi_1d\chi_1 \qquad 
\chi_1 = {1\over 2}(\Theta \l+ \bar{\Theta}\bar{\l}). 
\end{equation}
  From \p{chi1} we conclude that the odd momentum of $\chi$ is
proportional
to
$\chi$ itself 
\begin{equation}\plabel{S1}
S_{\chi} = \pi_{\chi_1} -4i\chi_1 =0.
\end{equation}
Eq. \p{S1} is the second--class constraint being
typical of any free fermion theory
\begin{equation}\plabel{S1S1}
\{ S_\chi, S_\chi \}_P = -8i.
\end{equation}
It can be regarded to be satisfied 
in the strong sense \cite{Dirac} after we pass 
from the Poisson brackets  to the  
Dirac brackets 
\begin{equation}\plabel{DB}
[ f, g\}_D 
= [ f, g\}_P 
-{i \over 8}[ f , S_\chi \}_P [ S_\chi, g \}_P 
\end{equation}
which imply
\begin{equation}\plabel{xixiD}
\{ \chi_1, \chi_1 \}_D = 2i.
\end{equation}
Hence, upon quantization $\xi$ becomes a Clifford
variable of odd Grassmann parity
\begin{equation}\label{cliff}
(\hat{\chi_1})^2=1.
\end{equation} 

The Clifford algebra generated by this variable consists of two
elements, the unit element and $\hat{\chi_1}$. Hence, all functions of 
$\hat{\chi_1}$ can be
written as a `Clifford algebra valued superfield' having two components
\cite{Sorokin87}
\begin{equation}\plabel{csf}
\Phi(\hat{\xi})=\phi + i\hat{\chi_1}\psi,
\end{equation}
where $\phi$ and $\psi$ do not depend on  $\hat{\chi_1}$.

We conclude that {\sl at $a=1$ the 
wave functions \p{twvfa<1}, \p{sPhi} become
Clifford `superfields' whose components again (as in the case
$a\not=0,1$) 
describe an infinite tower of fields of 
all possible helicities}.

We can decompose the  superfields \p{csf} 
  into the even and odd parts with respect to 
$\l$-parity and thus have the wave functions 
with definite Grassmann parity (bosonic and fermionic superfields). 

We see that at $a=1$ the model has the same spectrum of quantum
physical
states
as in the generic case.  
The difference with the generic case ($a\neq 1$) is  only in the 
transformation properties of the field components with respect to
target space supersymmetry -- the $a=1$  supersymmetry multiplets are
shortened (see  [1], Section 2).  
Note also that in the models with $a\ge 1$
one can impose additional reality condition on the quantum wave
functions.

In the Appendix we shall present another way of quantizing a single 
classical fermionic variable $\chi_{1}$ (see \p{chi1})
  which allows to treat it as 
a usual Grassmann variable and quantum superwave functions as
standard superfields.


\section{Quantization by using the conversion method}
\setcounter{equation}{0}

In order to justify the results of the supertwistor 
quantization of the model presented in Sect. 3 
and to clarify the space--time structure of the quantum 
wave functions, in this section we shall perform the quantization 
directly in the coordinate representation. 

Because of the appearance of a particular mixture of fermionic 
first and second class constraints there appears
a problem of quantizing the system covariantly. 
However, there exists a powerful method to handle this problem 
\cite{11,12,13}, which is based on  the conversion of the second class 
constraints into the first class ones.

The quantization of the Ferber-Shirafuji model by the conversion method 
was  considered in \cite{14,15,ES,17}. In \cite{ES} a  
$D=10$ supersymmetric particle with extra tensorial coordinates 
has been also discussed. In the present paper, however, 
the relation between spinor variables and the tensorial 
central charges, as well as their physical interpretation,
goes far beyond the results presented in \cite{ES}.

\bigskip

\subsection{Conversion degrees of freedom}

To convert the 
second class constraints into the first class ones we introduce   
additional (conversion) phase space 
degrees of freedom, whose number is equal to the the number of the 
second class  constraints.  

Thus, for the $a\not= 0,1 $ models  
we need {\bf 8}$_b$ $+$ {\bf 2}$_f$ conversion degrees of 
freedom. For this purpose we introduce
 bosonic spinors  $\rho_A$, $\bar{\rho}_{\dot{A}}$ 
plus its canonical momenta
$P_{(\rho )}^A$, $\bar{P}_{\bar{(\rho})}^{\dot{A}}$ 
\begin{equation}\plabel{brcvb}
[P_{(\rho )}^A, \rho_{B}]_P = \d_{B}^{~A}, \qquad 
[\bar{P}_{\bar{\rho}}^{\dot{A}}, \bar{\rho}_{\dot{B}}]_P = 
\d^{~\dot{A}}_{\dot{B}}, \qquad
\end{equation}
and two real fermionic variables $f_1$ and $f_2$ whose Poisson brackets
form a Clifford algebra
 \begin{equation}
 \plabel{SS1,2}
\{f_1, f_1 \}_P = -i (1-a),  \qquad 
\{f_1, f_2 \}_P = 0,  \qquad 
\{f_2, f_2 \}_P = -i (1+a).  \qquad 
\end{equation}
Instead of $f_1$ and $f_2$ we shall also use two conjugate
fermionic variables
 \begin{equation}
 \plabel{tS}
{S}= \sqrt{(1+a)} f_1 - i\sqrt{(1-a)}f_2  , \qquad
\bar{{S}}= \sqrt{(1+a)} f_1 + i\sqrt{(1-a)}f_2,  \qquad
\end{equation}
 \begin{equation}
 \plabel{tStS}
\{ {S},{S}\}_P = 0,  \qquad 
\{\bar{{S}}, \bar{{S}} \}_P = 0, \qquad 
\{ {{S}}, \bar{{S}} \}_P = -2i (1-a^2).
\end{equation}

Note,  that  
$\bar{{S}}$ is complex conjugate of ${{S}}$ only
for the case $0<a<1$, while for $a>1$, where $ \sqrt{(1-a)}= 
i\sqrt{|1-a|}$ is imaginary, both ${S}$ and $\bar{{S}}$ 
are independent real variables. 

For $a\not=1$
$\bar{{S}}$ can be regarded as the momentum conjugate to ${{S}}$.


\bigskip 

\subsection{Conversion of the constraints}

We use the additional degrees of freedom \p{brcvb} and \p{SS1,2}
in order 
to convert the mixture of  first and second class constraints 
\p{Phi1}--\p{bDA=0} into the first class ones.  

As it was shown in \cite{BMRS}, 
in twistor-like formulations of particle mechanics  
it is convenient to perform 
conversion  of the whole set of  primary constraints, without
dividing 
them into the sets of  first and second class constraints.

For any $a$ the first class constraints obtained as the result of
conversion 
are \footnote{We denote the converted constraints with the same letters
as the original ones.}
\begin{equation}\plabel{tPhi1}
{\Phi}_{A\dot{B}} \equiv
{P}_{A\dot{B}} - 
(\lambda_{A} + \rho_A ) (\bar{\lambda}_{\dot{B}} + \bar{\rho}_{\dot{B}}
)
= 0 ,
\end{equation}
\begin{equation}\plabel{tPhi2}
{\Phi}_{A{B}} \equiv
Z_{A{B}} - a (\lambda_{A} + \rho_A ) (\lambda_{B} + \rho_B ) = 0 ,
\end{equation}
\begin{equation}\plabel{tPhi3}
{\bar{\Phi}}_{\dot{A}\dot{B}} \equiv
\bar{Z}_{\dot{A}\dot{B}} - a 
(\bar{\lambda}_{\dot{A}} + \bar{\rho}_{\dot{A}} ) 
(\bar{\lambda}_{\dot{B}} + \bar{\rho}_{\dot{B}} )
= 0 ,
\end{equation}
\begin{equation}\plabel{tPA=0}
{{\Phi}}_{{A}}\equiv 
P^{A}_{(\lambda )} + P^{A}_{(\rho )} =0, \qquad
{\bar{\Phi}}_{\dot{A}}\equiv 
\bar{P}^{\dot{A}}_{(\bar{\lambda})}+\bar{P}^{\dot{A}}_{(\bar{\rho})} 
 = 0 ,
\end{equation}
\begin{equation}\plabel{tPrho=0} 
P^{A}_{(\lambda )}- P^{A}_{(\rho )} =0, \qquad
\bar{P}^{\dot{A}}_{(\bar{\lambda})} - \bar{P}^{\dot{A}}_{(\bar{\rho})} 
 = 0 ,
\end{equation}
 \begin{equation}\plabel{tDA=0}
{D}_{A}\equiv - \pi_A + i P_{A\dot{B}} \bar{\Theta}^{\dot{B}} + i
Z_{AB} \Theta^B  + (f_1+if_2) ~(\lambda_{A} + \rho_A ) 
=0, \qquad
\end{equation}
\begin{equation}\plabel{tbDA=0}
{\bar{D}}_{\dot{A}}\equiv  
\bar{\pi} _{\dot{A}} - i \Theta^B  P_{B\dot{A}}
- i \bar{Z}_{\dot{A}\dot{B}} \bar{\Theta}^{\dot{B}} + 
(f_1-if_2)~ (\bar{\lambda}_{\dot{A}} + \bar{\rho}_{\dot{A}} )= 0.
\end{equation}

The algebra of the first class constraints 
\p{tPhi1}--\p{tbDA=0} 
is quite simple. The only nonvanishing brackets (in the strong sense) 
appear in the fermionic sector and have the form 
\begin{equation}
 \label{DDD}
 {{\cal D}}_{\a\b} \equiv  \left( \matrix{
    \{{ D}_{A},{ D}_{B} \}_P  &  
\{ { D}_{A},{\bar{D}}_{\dot{B}} \}_P   \cr
     \{ {\bar{D}}_{\dot{A}},{D}_{{B}} \}_P &
 \{ {\bar{D}}_{\dot{A}},{\bar{D}}_{\dot{B}} \}_P \cr }\right)
  =  2i \left( \matrix{
    -{\Phi}_{AB} &  {{\Phi}}_{A\dot{B}}   \cr
    {{\Phi}}_{B\dot{A}}  &
 - { \bar{\Phi}}_{\dot{A}\dot{B}}  \cr }\right).
\end{equation}
The r.h.s. of \p{DDD} vanishes weakly, i.e. on the constraint surface 
\p{tPhi1}--\p{tPhi3}.

Note that the expressions \p{tPhi1}--\p{tPhi3}, \p{tDA=0} and \p{tbDA=0}
contain only the combination ($\l+\rho$) of the commuting spinors.
We denote this combination by $\tilde\l$
\begin{equation}
 \label{tlambda}
\tilde{{\l}}_{A}= \lambda_{A} + \rho_A , \qquad  
\tilde{\bar{\l}}_{\dot{A}} = 
\bar{\lambda}_{\dot{A}} + \bar{\rho}_{\dot{A}} ,
\end{equation}
while the linearly independent variables
\begin{equation}
 \label{trho}
\tilde{{\rho}}_{A}= \lambda_{A} - \rho_A, \qquad  
\tilde{\bar{\rho}}_{\dot{A}} = 
\bar{\lambda}_{\dot{A}} - \bar{\rho}_{\dot{A}}
\end{equation}
completely decouple and
have vanishing canonical momenta \p{tPrho=0}. Hence,  the variables
\p{trho}
can be excluded from the consideration, since the wave functions
will not depend on these variables.

\subsection{Quantization of the converted system: 
equations for the wave function}

Now it is straightforward to quantize the system by using the Dirac
method  
\cite{Dirac}. 
For this purpose  let us choose the (super-)Shr\"{o}dinger 
representation for the superspace coordinates and the bosonic spinor 
variables  
\begin{equation}\plabel{hatP4D}
\hat{P}_{A\dot{B}} = - i 
{\partial \over \partial X^{A\dot{B}}}, \qquad 
\hat{Z}_{A{B}} = - { i} 
{\partial \over \partial y^{A{B}}}, \qquad 
\hat{\bar{Z}}_{\dot{A}\dot{B}} = - i 
{\partial \over \partial \bar{y}^{\dot{A}\dot{B}} }, \qquad 
\end{equation}
$$
\hat{P}^{A}_{\tilde{\l} }= - i 
{\partial \over \partial \tilde{\l}^{A}}, 
\qquad 
\hat{\bar{P}}^{\dot{A}}_{\tilde{\bar{\l}} }= - i 
{\partial \over \partial \tilde{\bar{\l}}^{\dot{A}}}, 
\qquad 
$$
\begin{equation}\plabel{hatpi4D}
\hat{\pi}_{A} = + i 
{\partial \over \partial \Theta^{A}},  
\qquad \hat{\bar{\pi}}_{\dot{A}} = + i 
{\partial \over \partial \bar{\Theta}^{\dot{A}}}.
\qquad 
\end{equation}
The fermionic variables $f_1$ and $f_2$ become Clifford algebra
operators
\begin{equation}
\plabel{brcvfq}
({\hat{f}_1})^2 = {1\over 2}(1-a) , \qquad 
({\hat{f}_2})^2 = {1\over 2}(1+a) , \qquad 
\{\hat{f}_1,\hat{f}_2  \} = 0. \qquad 
\end{equation}
The Grassmann parity of $f_1$ and $f_2$ must 
be odd because the constraints \p{tDA=0} and \p{tbDA=0} should
  have definite parity.

Note that the linear combinations \p{tS} of fermionic quantum variables
  $f_1$ and $f_2$ 
satisfy the commutation relations 
\begin{equation} \plabel{qtStS}
\{ \hat{S}, \hat{S} \} = 0,  \qquad 
\{ \bar{\hat{S}}, \bar{\hat{S}} \} = 0, \qquad 
\{ {\hat{S}}, \bar{\hat{S}} \} = 2 (1-a^2).
\end{equation}
So one can choose $S$ as an odd coordinate and $\bar{\hat{S}}$ as 
its momentum operator 
$$\hat{\bar{S}}=2(1-a^2){\partial\over{\partial {S}}}, \qquad
\hat{S}=S$$

Despite of the fact that such a representation makes hermiticity
condition nonmanifest, 
it is convenient since it simplifies the calculations and 
provides the possibility of treating the cases  $0<a<1$ and $a>1$ 
on an equal footing.

\bigskip 

After quantization the first--class constraints \p{tPhi1}-\p{tPA=0},
\p{tDA=0} and \p{tbDA=0}
are imposed on  the wave function  
\begin{equation}
\plabel{wf4D}
 \Psi = \Psi (x^{A\dot{A}}; y^{AB}, \bar{y}^{\dot{A}\dot{B}};
\tilde{\l}^A, \tilde{\bar{\l}}^{\dot{A}}; 
\Theta^A, \bar{\Theta}^{\dot{A}}, S)
\end{equation}
(we recall that we have consistently removed the variables \p{trho} from
the consideration).

Thus, the wave function of the system satisfies the first 
order differential equations
\begin{equation}\plabel{tPhi1q}
({\partial \over \partial x^{A\dot{B}}} - i 
\tilde{\lambda}_{A} \tilde{\bar{\lambda}}_{\dot{B}} )
\Psi = 0 ,
\end{equation}
\begin{equation}\plabel{tPhi2q}
({1 \over 2}
{\partial \over \partial y^{A{B}}} -
ia\tilde{\lambda}_{A}\tilde{\lambda}_{B})
\Psi = 0 ,
\end{equation}
\begin{equation}\plabel{tPhi3q}
({1 \over 2}
{\partial \over \partial 
\bar{y}^{\dot{A}\dot{B}}} - i a \tilde{\bar{\lambda}}_{\dot{A}}
 \tilde{\bar{\lambda}}_{\dot{B}})
\Psi = 0 ,
\end{equation}
\begin{equation}\plabel{tDA=0q}
\left({\partial \over \partial \Theta^A} + i
 {\partial \over \partial x^{A\dot{B}}} \bar{\Theta}^{\dot{B}} 
+ {i \over 2} a {\partial \over \partial y^{A{B}}}\Theta^B 
- i (f_1+if_2) \tilde{\lambda}_{A}\right)
\Psi = 0 ,
\end{equation}
\begin{equation}\plabel{tbDA=0q}
\left({\partial \over \partial \bar{\Theta}^{\dot{A}}} 
+ i {\partial \over \partial x^{B\dot{A}}} \Theta^B 
+ {i \over 2} a {\partial \over \partial \bar{y}^{\dot{A}\dot{B}}} 
\bar{\Theta}^{\dot{B}} 
+i (f_1-if_2)\bar{\lambda}_{\dot{A}}\right)
\Psi = 0 , 
\end{equation}

The solution of eqs. \p{tPhi1q}--\p{tPhi3q} is  
\begin{equation}
\plabel{wf4Ds1}
 \Psi = e^{
i \l_A \bar{\lambda}_{\dot{A}} x^{A\dot{A}}
+ ia \l_A\l_By^{AB} + ia \bar{\lambda}_{\dot{A}}\bar{\lambda}_{\dot{B}}
\bar{y}^{\dot{A}\dot{B}}}
~~~
g( 
\tilde{\l}^A, \tilde{\bar{\l}}^{\dot{A}}; 
\Theta^A, \bar{\Theta}^{\dot{A}}, S). 
\end{equation}

Because of the constraints \p{tDA=0q} and  \p{tbDA=0q}
the function $g$ \p{wf4Ds1} satisfies the conditions
\begin{equation}\plabel{tDA=0q1}
\left({\partial \over \partial \Theta^A} - 
\tilde{\lambda}_{A} [(\bar{\Theta}\bar{\lambda}) 
- a ({\Theta}{\lambda}) - i(f_1+if_2) ]\right)
g = 0 ,
\end{equation}
\begin{equation}\plabel{tbDA=0q1}
\left({\partial \over \partial \bar{\Theta}^{\dot{A}}} 
- \bar{\lambda}_{\dot{A}} [({\Theta}{\lambda}) - 
a (\bar{\Theta}\bar{\lambda}) 
+ i (f_1-if_2)]\right) g = 0 , 
\end{equation}

An evident consequence of eqs. \p{tDA=0q1} and \p{tbDA=0q1}
is that $g$ depends only on the composite Grassmann variables
  $\chi=\Theta^B\l_B$ and 
$\bar\chi= \bar{\Theta}^{\dot{B}}\bar{\l}_{\dot{B}}$ introduced in
\p{Pcorr(a)f}
\begin{equation}
\plabel{wf4Ds2}
g( 
\tilde{\l}^A, \tilde{\bar{\l}}^{\dot{A}}; 
\Theta^A , \bar{\Theta}^{\dot{A}}, S)=
g( 
\tilde{\l}^A, \tilde{\bar{\l}}^{\dot{A}}; 
\Theta^B\l_B , \bar{\Theta}^{\dot{B}}\bar{\l}_{\dot{B}}, S). 
\end{equation}
Then the eqs. \p{tDA=0q1} and \p{tbDA=0q1} 
reduce to
\begin{equation}\plabel{tDA=0q2} 
\left({\partial \over \partial \chi} - \bar{\chi } - a 
\chi - i ({f_1}+if_2) \right) 
~g(\l_A, \bar{\l}_{\dot{A}}; \chi , 
\bar{\chi }, S) = 0 ,
\end{equation}
\begin{equation}\plabel{tbDA=0q2}
\left({\partial \over {\partial \bar{\chi}}} - {\chi}- a\bar{\chi} 
 + i ({f_1}-if_2) \right) 
~g (\l_A, \bar{\l}_{\dot{A}}; \chi , 
\bar{\chi }; S) = 0.
\end{equation}

\subsection{Quantization of the converted system: 
 dependence on the fermionic variables}

To find the solution of equations \p{tDA=0q2} and \p{tbDA=0q2}
we take their linear combinations and rewrite them
in the following form
\begin{equation}\plabel{dchiS}
\left[\sqrt{1-a^2}({\partial\over{\partial\bar\chi(a)}}
-\chi(a))-S\right]g=0,
\end{equation}
\begin{equation}\plabel{dchibS}
\left[\sqrt{1-a^2}({\partial\over{\partial{\chi(a)}}}-\bar\chi(a))-2(1-a^2)
{\partial\over{\partial S}}\right]g=0,
\end{equation}
where $\chi(a)$ were introduced in \p{chi}, \p{chia} and $S$ are defined
in \p{tS}.

The equations \p{dchiS} and \p{dchibS} are easily solved in terms of the
components of the superfunction $g(S)$
\begin{equation}\plabel{gs}
g(S)=g_0(\chi,\bar\chi)+iSg_1(\chi,\bar\chi)
\end{equation}
which satisfies the conditions
\begin{equation}\plabel{g0}
({\partial\over{\partial\bar\chi(a)}}
-\chi(a))g_0=0\quad \Rightarrow
g_0(\chi,\bar\chi)=e^{-\bar\chi(a)\chi(a)}
\Phi(\l_A, \bar{\l}_{\dot{A}}; \chi(a)),
\end{equation} 
\begin{equation}\label{g1}
2\sqrt{1-a^2}g_1(\chi,\bar\chi)=
-i({\partial\over{\partial{\chi(a)}}}-\bar\chi(a))g_0.
\end{equation}
We see that when $a\not =1$ the component $g_1$ of the superfunction
\p{gs} is expressed in terms of $g_0$ which is  specified by
the condition \p{g0} in terms of a single independent superfield
$\Phi(\l_A, \bar{\l}_{\dot{A}}; \chi(a))$.

Hence, the independent wave function \p{wf4D} which describes 
the general solution of the equations \p{tPhi1q}--\p{tbDA=0q} is
\begin{equation}
\plabel{wf4Ds3}
 \Psi = e^{
i \l_A \bar{\lambda}_{\dot{A}} x^{A\dot{A}}
+ ia \l_A\l_By^{AB} + ia \bar{\lambda}_{\dot{A}}\bar{\lambda}_{\dot{B}}
\bar{y}^{\dot{A}\dot{B}} - \bar\chi(a)\chi(a)}
~~~
\Phi( 
\tilde{\l}^A, \tilde{\bar{\l}}^{\dot{A}};\chi(a)). 
\end{equation}
At the critical value $a=1$ the result is the same though the proof is
slightly changed since in such a case  
(as in Subsection 3.1.4)  we should deal with a single 
conversion Clifford variable
$f$ instead of $f_1$ and $f_2$ (and/or $S$ and $\bar S$). 
More precisely, in eqs. \p{tDA=0q2} and \p{tbDA=0q2} one should put a=1,
$\bar\chi=\chi$ and $f_1=S=0$, and then follow the quantization
prescription
described in the Appendix.

One should notice that the wave function 
$\Phi(\tilde{\l}^A, \tilde{\bar{\l}}^{\dot{A}};\chi(a))$  
in \p{wf4Ds3} has exactly
the same structure as in the supertwistor case \p{twvfa<1}, but where
now $\chi(a)$  
are the composite Grassmann coordinates
defined by eqs. \p{chi}, \p{chia}. 
We therefore conclude that the direct supertwistor 
quantization and the quantization with the use of conversion 
of the superparticle model
based on the generic action \p{ac(a)}
result
in  the same supersymmetric spectrum of 
the quantum states.

The supersymmetry transformations of the components of $\Phi( 
\tilde{\l}^A, \tilde{\bar{\l}}^{\dot{A}};\chi(a))$ are easily derived
from
\p{wf4Ds3} using the supersymmetric variations \p{susy} of the
coordinates.

The higher dimensional 
generalization  of the $a=1$ model and its quantization 
will be the subject of the next section. 

\section{The $a=1$ model in higher dimensions and internal
degrees of freedom}
\setcounter{equation}{0}

A  generalization of the $D=4$, $a=1$ superparticle model has been 
proposed in \cite{BL}. 
In higher space--time dimensions $D$ we consider an extension of an 
$N=1$ supersymmetry algebra by tensorial central charges
\begin{equation}\plabel{n1cc}
\{Q_\a,Q_\b\}=P_{\a\b}, \quad [P_{\a\b},Q_\g]=0,
\end{equation}
where, depending on space--time dimension $D$, the supercharges $Q_\a$ 
$(\a = 1,\ldots 2^k)$
are real Majorana, or Majorana--Weyl
  spinors\footnote{\label{lufoot} In the general case one can also
consider  the 
cases of pseudo--Majorana, simplectic--Majorana  and Dirac (complex)
supercharges \cite{west}. 
 Technical details of the extension of the results of
Section 5 to arbitrary type of supercharges will be considered
in another publication.} and $P_{\a\b}$ is
a symmetric generalized `momentum' generator conjugate to
$2^k (2^k +1)$ symmetric spin-tensor coordinates $X^{\a\b}$, which can
be 
split into the usual space--time coordinates and tensorial 
central charge coordinates, as we shall demonstrate below.

We assume that $P_{\a\b}$ is defined by the Cartan--Penrose 
relation 
 \begin{equation}\plabel{CPrepD}
 P_{\a\b} = \l_{\a} \l_{\b}, 
 \end{equation}
where the real bosonic spinor $\l_{\a}$ has the same spinor properties
as
the supercharge $Q_\a$.

The expression \p{CPrepD} implies the BPS condition
$det P_{\a\b}=0$ and can be obtained as a primary constraint 
from the action functional  \cite{BL}
\begin{equation}\plabel{actionD}
S = \int_{{\cal M}^1} \l_{\a} \l_{\b} \Pi^{\a\b}
\end{equation}
$$
 \Pi^{\a\b} = dX^{\a\b} - i d\Theta^{(\a}\Theta^{\b )}
=d\tau \Pi^{\a\b}_\tau , \qquad 
$$
$$
\a =1,...,2^k.
$$

\bigskip 

For any value of $k$ the model possesses 
$2^k$ global target space supersymmetries generated by \p{n1cc}
$$
\d_{susy} X^{\a\b} = i \Theta^{(\a}\e^{\b )}, \qquad 
\d_{susy} \Theta^{\a}=\e^{\a }, \qquad 
\d_{susy} \l_{\a} = 0, \qquad 
$$
as well as  $2^k-1$ $\kappa$--symmetries. 

To show the presence of the  $2^k-1$  $\kappa$--symmetries let us 
write the variation of the action \p{actionD}
\begin{equation}\plabel{actionDv}
\d S = \int_{{\cal M}^1} 
\left(2 \d \l_{\a} ~\l_{\b} \Pi^{\a\b} 
+ d (\l_{\a} \l_{\b}) i_\d \Pi^{\a\b} - 
2i d\Theta^\a \l_{\a} \l_{\b}\d\Theta^\b
\right) + 
\left(\l_{\a} \l_{\b} i_\d \Pi^{\a\b} \right)\vert_{\tau_i}^{\tau_f},
\end{equation}
where
$$
 i_\d \Pi^{\a\b} = \d X^{\a\b} - i \d \Theta^{(\a}\Theta^{\b )}. \qquad 
$$

One can see that only one linear combination   $\l_{\b}\d\Theta^\b$ 
of $2^k$ independent variations of Grassmann coordinates 
$\d\Theta^\b$ is effectively involved into the variation of the action. 

Hence, other $2^k-1$ Grassmann coordinate variations (which do not
appear in
\p{actionDv}) can be identified with 
the parameters of local fermionic $\kappa$--symmetry. They
can be written in the form 
$$\kappa^I= u^I_{\b}\d\Theta^\b, $$  
where 
$u^I_{\b}$ ($I=1,\ldots, 2^k-1$) and $\l_\b$ 
form a set of $2^k$ linearly independent  bosonic  
spinors.

\bigskip
Identifying the $\kappa$ symmetry with the part of  
target space supersymmetry which is 
preserved by the particle or brane configuration, 
we  claim that the model \p{actionD} describes the dynamics of 
BPS states preserving all but one target space supersymmetries 
in space--time of a dimension $D$.

\bigskip
{\bf Examples.}

\bigskip
In $D=3$ (where $k=1$) the action \p{actionD}  describes the 
standard massless superparticle
$$ 
k=1~\leftrightarrow ~D=3: \quad  X^{\a\b} = X^m \g_m^{\a\b},
\quad  P^{\a\b} = P^m \g_m^{\a\b}.
$$
On the other hand the case $k=1$ can be regarded as a model in a
`minimal' 
$D=2+2$ superspace with self-dual tensorial central charge coordinates 
$X^{\a\b}= y^{mn}\s_{mn}^{\a\b}$, $y^{mn} = {1\over 2} \e^{mnkl}
y_{kl}$.

\bigskip 

The case of $k=2$  corresponds to the $D=4$, $a=1$ model considered
in Sections 2--4
but written in the Majorana representation. 

\bigskip
The construction also holds in $D=6$ where $k=3$,  but here
 we should use the ($SU(2)$--Majorana--Weyl) `reality' conditions.  
  In addition to the 4--dimensional spinor index
$\a$ the complex 8--component 
  spinors $Q_\a^i$ and $\l_\a^i$ carry the $SU(2)$ index $i=1,2$ 
and they are the $SU(2)$ Majorana--Weyl spinors 
(see for details \cite{west}). 
The number of tensorial central
charges in this model is $30$.

The case $k=4$ can be regarded as describing a 
$D=10$ massless superparticle with $126$ composite 
(self--dual) tensorial central charges
$Z_{m_1 \ldots m_5}$ \cite{BL} (cf. with \cite{HP,ES}). 
The real supercharges $Q_{\alpha}$ satisfy the Majorana--Weyl
reality condition.

\bigskip

The action \p{actionD} with $k=5$ corresponds to a
$0$--superbrane model
in $D=11$ superspace with
$517$ tensorial central charges
composed from $32$ components of one real bosonic Majorana 
spinor. In contrast to the cases of $D=3,4,6$ and 10, in such a model
the  superparticle is not massless, the mass of the 0-brane 
being generated dynamically 
in a way similar to the mechanism generating the tension of 
superstrings and superbranes \cite{generation} 
(see \cite{BL} for some details).

\bigskip 

On the other hand it is possible to use the 
twelve dimensional $D=2+10$ $32\times 32$ gamma matrices 
to treat the $k=5$ model from the point of view 
of two--time physics \cite{Bars}. 
The bosonic coordinates $X^{\a\b}$ 
are decomposed into two--index and self--dual 6--index 
central charge coordinates $y^{mn}, y^{m_1\ldots m_6}= 
1/6! \epsilon^{m_1\ldots m_6 n_1\ldots n_6} y_{n_1\ldots n_6}$. 

\bigskip 

\subsection{$OSp(1|2^k)$ supertwistor representation of the 
D-dimensional model}

Performing the integration by parts we can rewrite the action
\p{actionD}
in the $OSp(1|2^k)$ invariant form (i.e. $OSp(1|16)$ for $D=10$ 
and $OSp(1|32)$
for $D=11$) in terms of a supertwistor $Y^{{\cal A}}= (\mu^\a, \zeta )$ 
\begin{equation}\plabel{actiontwD}
S= - \int (\mu^\a d\l_\a + i d\zeta ~\zeta ), \qquad \a = 1,\ldots ,
2^k.
\end{equation}
The generalized Penrose--Ferber  correspondence
between  the supertwistors and  the generalized superspace
looks as follows
\begin{equation}\plabel{muXZD}
 P_{\a\b} = \l_{\a} \l_{\b},\quad 
\mu^\a = X^{\a\b} \l_\b - i \Theta^\a (\Theta^\b \l_\b), \quad
\zeta = \Theta^\a \l_\a .
\end{equation}
and does not imply other constraints. 

\subsection{Quantization of the higher dimensional model 
with the use of conversion}

The  quantization of the supertwistor formulation 
\p{actionD} is straightforward and is completely analogous to 
the quantization of the $D=4$ model considered in Subsection {\bf 3.1.4} 
and Appendix. 

The spectrum of quantum states is  described by the  
superfield 
\begin{equation}\plabel{wftwD}
\Phi = \Phi 
( \l_\a, \zeta ) = 
\phi ( \l_\a) + i  \zeta \psi  ( \l_\a) 
\end{equation}
depending on the bosonic spinor 
$\l_\a$ and one Grassmann (or, equivalently, Clifford) variable $\zeta
$.

\bigskip

For completeness we briefly describe the quantization of the 
higher dimensional model \p{actionD} with the use of conversion.

The primary constraints of the model \p{actionD} 
are 
\begin{equation}\plabel{Phi}
\Phi_{\a\b} \equiv P_{\a\b} - \l_\a\l_\b = 0
\end{equation}
\begin{equation}\plabel{D}
D_{\a} \equiv \pi_{\a} + i \Th^\b P_{\b\a} = 0
\end{equation}
\begin{equation}\plabel{Pl}
P^{\a}_{(\l )}  = 0
\end{equation}
where the momenta are defined in such a way that the  
nonvanishing Poisson brackets have the following form 
\begin{equation}\plabel{Pbr}
[P_{\a\b}, X^{\g\d}]_P= 2\d_{(\a}^{(\g}\d_{\b )}^{\d) } , \qquad 
\{ \pi_{\a}, \Th^{\b} \}_P = \d_{\a}^{\b} , \qquad 
[P^{\a}_{(\l )}, \l_\b ]_P = \d^{\a}_{\b} . \qquad 
\end{equation}

This set of  $2^k (2^k + 1) / 2$ bosonic and $2^k$ fermionic constraints 
obeys the algebra 
\begin{equation}\plabel{algD}
[\Phi_{\a\b}, P^{\g}_{(\l )}]_P= 2\l_{(\a} \d_{\b )}^{\d} , \qquad 
\{ D_{\a}, D_{\b} \}_P = 2i P_{\a\b}
\equiv 2i (\Phi_{\a\b} + \l_\a\l_\b),
\end{equation}
$$
all~other~brackets~= 0, 
$$
and thus contains  $2^k$ bosonic and $1$ fermionic second class
constraints. 
Therefore, our system with 
 $2^k (2^k + 1) / 2$ bosonic and $2^k$ fermionic configuration space 
variables contain  $2^k$ bosonic and $1$ fermionic
 physical degrees of freedom which can be identified with the
components of 
$OSp(1|2^k)$ supertwistor.

\bigskip 

Exactly as in the $D=4$ case,
to perform the conversion (see \cite{9}--\cite{13}) 
of the second class 
constraints into the first class ones we introduce  
additional 'conversion' degrees of freedom (two for each 
pair of the bosonic second class constraints and one 
self--conjugate fermionic variable for each fermionic 
second class constraint) 
\begin{equation}\plabel{Pbrmu}
[P^{\a}_{(\rho )}, \rho_\b ]_P = \d^{\a}_{~\b}, \qquad 
\{ \xi , \xi \} = - {i \over 2}, \qquad 
\end{equation}
and transform the second--class constraints into first--class ones 
extending the former with the new coordinates and momenta 
\begin{equation}\plabel{tPhi}
\tilde{\Phi}_{\a\b} \equiv P_{\a\b} - \tilde{\l}_\a \tilde{\l}_\b= 0 , 
\end{equation}
\begin{equation}\plabel{tD}
\tilde{D}_{\a} \equiv D_\a + 2 \tilde{\l}_\a \xi \equiv  
\pi_{\a} + i \Th^\b P_{\b\a} + 2
  \tilde{\l}_\a  
  \xi = 0, 
\end{equation}
\begin{equation}\plabel{tPl}
P^{\a}_{\tilde{\rho} } = 0.
\end{equation}
where
\begin{equation}\plabel{tildel}
\tilde{\l}_{\a}=\l_\a + \rho_{\a},  \qquad \tilde{\rho}_{\a}=\l_\a -
\rho_{\a}.
\end{equation}

Following the  
Appendix we obtain the superwave function describing the 
first--quantized states of the model determined
by the single superfield \p{wftwD} 
depending on  
$\tilde{\l}_\a$ and one Grassmann variable $\chi= (\Th\l)$.
We have

\begin{equation}\plabel{Psisol}
 \Psi(\tilde{\l}_a, (\Th\tilde\l))= 
e^{{i \over 2} \tilde{\l}_\a\tilde{\l}_\b X^{\a \b}}
\left[\phi(\tilde{\l}_a) 
+ i(\Th\tilde\l )~  \psi (\tilde{\l}_a)\right].  
\end{equation}

In the sector with 
even $\lambda$--parity of the wave function ($\Psi=\Psi_{+}$)  
  the spectrum of the quantum states of the model 
\p{actionD} is described by one bosonic  $\phi_+(\tilde{\l}_a)$
and one fermionic $ \psi_-(\tilde{\l}_a)$ function, 
while in the $\l$--odd sector ($\Psi=\Psi_{-}$) we have the
fermionic field $\phi_-(\tilde{\l}_a)$ and the bosonic field 
$ \psi_+(\tilde{\l}_a)$. 
This  is in complete correspondence with the result of the
quantization  of the free supertwistor  model  \p{actiontwD}.

\subsection{Properties of the wave function with arbitrary
helicity spectrum}

To clarify the meaning of the wave function \p{wftwD} (or  
\p{Psisol}), 
let us consider 
its bosonic limit at
$\Theta^\a=0$
\begin{equation}\plabel{Psisol0}
 \Psi= e^{{i \over 2} \tilde{\l}_\a\tilde{\l}_\b X^{\a \b}}
\phi(\tilde{\l}_\a), \qquad \a=1,\ldots , 2^k,    
\end{equation}
and use the decomposition of the product of 
the spinor representations in the basis of $D$-dimensional
gamma-matrices. 

\bigskip 

For the simplest case  $k=1$ ($\a =1,2$), where our model coincides with 
a $D=3$ counterpart of the usual (Ferber-Shirafuji) model
\cite{F78,S83}, 
the Fierz identity reads
\footnote{
We use the matrices
$\gamma^{a}_{\alpha \beta}$ which are symmetric and obtained    
 from standard Dirac matrices 
$(\gamma^{a})_{\alpha}^{~\beta}$ by lowering one of the indices with the 
charge conjugation matrix
 $C=\gamma_{0}= i \tau_2$.} 
$$
D=3: \qquad \l_\a \l_\b = {1 \over 2} \gamma^a_{\a\b} (\l \g_a\l)=
{1 \over 2} \gamma^a_{\a\b} P_a,  
$$ 
and we can identify the matrix coordinates $X^{\a\b}$ 
and their momenta  $P_{\a\b}$ with the usual vector
coordinates and momenta 
$$
D=3: \quad X^a = {1 \over 2} \gamma^a_{\a\b}X^{\a\b}, \quad     
X^{\a\b}= X^a  \gamma_{a\a\b}; \quad 
P^a = {1 \over 2} \gamma^a_{\a\b}P^{\a\b}, \quad     
P_{\a\b}= P^a  \gamma_{a\a\b}. \qquad     
$$ 
Thus, in $D=3$ eq. \p{Psisol0} describes  a plane wave solution\footnote{More precisely, in $D=3$  $\phi( \l_a ) = 
\psi (p_m, sign(\l ))$, where $sign(\l )$ denotes the 
sign factor $(\pm 1)$ of the bosonic spinor. 
This is a `parameter' of the residual {\bf Z}$_2$ 
symmetry, whose action on $\l$ does not change 
$p_m$.}  

\begin{equation}\plabel{Psisol3D0}
D=3: \qquad  \Psi= e^{i p_m X^{m}}\phi (p_m).
\end{equation}

\bigskip 

The case $k=2$, $D=4$ has been analized in detail in Sections 2--4.
To transform the wave function \p{Psisol0} to the wave function
\p{wf4Ds3} (at $a=1$, and $\Theta=0$) one should perform the similarity
transformation from the real Majorana  to the
complex Weyl representation of the $D=4$ gamma--matrices 
and replace the  Majorana spinor by the pair of complex conjugate Weyl
spinors 
\begin{equation}\plabel{sdec4D}
\tilde\l_\a ~~ \leftrightarrow 
\pmatrix{  \tilde{\l}_A  \cr
{\bar{\tilde{\lambda}}}_{\dot{A} }\cr
}.
\end{equation}

 In the momentum representation 
the wave function $\phi (\l_\a )$ differs
from the usual one given by  
 $\phi_0 (p_m)$ by the presence of additional dependence on the 
angle 
variable  $\a$ which describes the common phase factor
 of the Weyl spinor $\lambda_{A}$  $(\lambda_{1} = e^{i(\alpha
+\beta)} |\lambda_{1}|$, $\lambda_{2}=e^{i(\alpha - \beta)}
|\lambda_{2}|$) and  parametrizes the 1-dimensional
sphere $S^1$ 
$$
D=4: \qquad \phi (\l_a)=\phi (\l_A, \bar{\l}_{\dot{B}})
= \phi (p_m, \a), \qquad \a \in [0,2\pi)
$$
where $
p_m = {1 \over 4} \l \g_m \l = {1 \over 2} \l^A {\s}_{A\dot{A}} 
\bar{\l}^{\dot{A}}$ (see also $^{10}$). 
The additional internal 
 momentum variable  $\a$ 
is the  only independent  degree of freedom  
contained in the  $D=4$ tensorial central charges 
composed of the bosonic spinor 
$$ \a  \in [0,2\pi) 
\qquad \Leftrightarrow \qquad Z_{mn} = {1 \over 4} \l\g_{mn} \l
\qquad 
$$
It describes the $D=4$ helicity spectrum  
of the quantum states.

\bigskip 

In the general case of $k > 2$ with $2^k$ equal to the dimension 
of an irreducible spinor representation of $SO(1,D-1)$  in 
$D= 3,4,6, ~10 ~(mod~8)$ (i.e. $k= 3,4,\ldots$) the discussion
is similar. 
For example, in the case $k=4$, $D=10$ we can use the basis 
of symmetric $\sigma$ matrices
$\s_m, \s_{m_1 \ldots m_5}$ to make the decomposition 
\begin{equation}\plabel{ll10D}
\l_{\a} \l_{\b} \equiv P_{\a\b} =
P_m \s^m_{\a\b}+
Z_{m_1...m_5}
\s^{m_1...m_5}_{\a\b} ,
  \end{equation}
where 
\begin{equation}\plabel{P10D}
P_m = {1 \over 16} \l_{\a} \s_m^{\a\b}\l_{\b}    \qquad
\Rightarrow \qquad P_mP^m = 0
\end{equation}
is an ordinary  light--like momentum vector in $D=10$ and 
\begin{equation}\plabel{Z10D}
Z_{m_1...m_5} = {1 \over 16\cdot 5! } \l_{\a} \s_{m_1...m_5}^{\a\b}\l_{\b}   
\qquad
\end{equation}
is the momenta canonically conjugate to the {\bf 126} 
tensorial central charge coordinates 
$y^{m_1...m_5}$. 

It was demonstrated that 
the $D=10$ model contains the local
symmetries 
 (first class constraints) 
and second--class constraints which reduce the number of the classical 
bosonic degrees 
of freedom to the ones described by the 
 $16$--component bosonic spinor
$\l_\a$ and its momentum plus one Grassmann degree of freedom.  
In the quantum theory this is reflected in the dependence of 
the `momentum space representation' of the wave function
on 16 bosonic spinor variables and one Grassmann variable only. 

Due to the identities $(\s_m)_{(\a\b} (\s^m)_{\g ) \d}=0$ the 
$D=10$ momentum \p{P10D} is light--like. 
Hence, the 
tensorial central 
charge momenta  $Z_{m_1...m_5}$ contain $16-9= 7$ additional
degrees of freedom  which are not determined by the 
light-like momentum. 

\bigskip 

We now show that these additional internal
  degrees of freedom parametrize an $S^7$ sphere.
For this purpose we 
 perform a Lorentz transformation to a frame where the 
light--like momentum \p{P10D} acquires the form 
\begin{equation}\plabel{P10Dst}
P_m= (p, 0,0,0,0,0,0,0,0, p).
\end{equation}
Then in this frame we make an $SO(8)$ invariant split of the bosonic
spinor 
$\tilde{\l}_\a$ 
\begin{equation}\plabel{sdec10D}
\tilde{\l}_\a =\left(\matrix{\L_q \cr \S_{\dot{q}}\cr }\right), \qquad 
q=1, \ldots 8, \qquad \dot{q} = 1, \ldots 8
\end{equation} 
 and choose the $SO(8)\times SO(1,1)$ covariant representation
for 
the D=10 $\sigma$-matrices
$$
\s^{\underline{ 0}}_{\underline{\a}\underline{\b}}=
\hbox{ {\it diag}}(\delta _{{ qp}},
\delta_{{\dot q}{\dot p}})
= \tilde{\s }^{\underline{0}~\underline{\a}\underline{\b}} ,
\qquad \s^{\underline{9}}_{\underline{\a}\underline{\b}}=
\hbox{ {\it diag}} (\delta _{qp},
-\delta _{{\dot q}{\dot p}}) =
-\tilde{\s }^{\underline{9}~\underline{\a}\underline{\b}} ,
$$
\begin{equation}\plabel{gammarep}
\s^{i}_{\underline{\a}\underline{\b}} =
\left(\matrix{0 & \gamma ^{i}_{q\dot p}\cr
\tilde{\gamma}^{i}_{{\dot q} p} & 0\cr}
\right)
= - \tilde{\s }^{i~\underline{\a}\underline{\b}} ,
\end{equation}
$$
\s^{{++}}_{\underline{\a}\underline{\b}} \equiv
(\s^{\underline{ 0}}+
\s^{\underline{ 9}})_{\underline{\a}\underline{\b}}=
\hbox{ {\it diag}}(~2\delta _{qp},~ 0)
= -(\tilde{\s }^{\underline{ 0}}-
\tilde{\s }^{\underline{ 9}})^{\underline{\a}\underline{\b}} =
\tilde{\s }^{{--}~\underline{\a}\underline{\b}} ,
\qquad 
$$ 
$$
\s ^{{--}}_{\underline{\a}\underline{\b}}\equiv
(\s ^{\underline{ 0}}-\G ^{\underline{ 9}}
)_{\underline{\a}\underline{\b}}=
\hbox{ {\it diag}}(~0, ~2\delta
_{{\dot q}{\dot p}}) = (\tilde{\s }^{\underline{ 0}}+
\tilde{\s }^{\underline{ 9}})^{\underline{\a}\underline{\b}}
= \tilde{\s }^{{++}~\underline{\a}\underline{\b}}.
$$
In the frame \p{P10Dst} the Cartan-Penrose representation \p{P10D} 
 looks as follows
\begin{equation}\plabel{CP10Dst}
\L_q~\L_q = p, \qquad 
 2 \S_{\dot q} \S_{\dot q} =0, \qquad
 \L_{q} \g^i_{q\dot{p}} \S_{\dot{p}}=0 
\end{equation}
The general solution of eqs. \p{CP10Dst}
is 
\begin{equation}\plabel{sdec10Dsol}
\S_{\dot q} =0, \qquad \Rightarrow \qquad \tilde{\l}_\a
=\left(\matrix{\L_q \cr 0\cr }\right), \qquad 
\end{equation}  
and the only nonvanishing component of the momentum  \p{P10Dst}
is given by the norm of the $SO(8)$ spinor $\L_q$
\begin{equation}\plabel{CP10Dst0}
p=\L_q~\L_q . \qquad 
\end{equation}
The expression \p{CP10Dst0} is invariant under the $SO(8)$ rotations
$$
\L_q ~\rightarrow~\L_p S_{pq}, \qquad S~S^T=I. 
$$ 
But not all $SO(8)$ transformations act  on  $\L_q$ effectively. 
Indeed, if one fixes the $SO(8)$ gauge 
\begin{equation}\plabel{sga10D}
\L_q = \left(\matrix{ \pm \sqrt{p} \cr 0 \cr \ldots \cr  \ldots \cr
\ldots \cr 
                                                   0 \cr }\right)
\end{equation} 
one  finds that i) this gauge is invariant under the $SO(7)$
transformations, 
ii) any form of the spinor $\L_q$ can be obtained from
\p{sga10D} 
by a transformation from the coset space 
$SO(8)/SO(7)$ isomorphic to the sphere $S^7$. 

\bigskip

Thus, the 16 components of the bosonic spinor 
$\l_\a$ in $D=10$ can be split into (double covering ($\lambda
\simeq - \lambda$)) 
\\ i)  degrees of freedom which
 characterize the light-like momentum $P_m$, 
\\ ii) 7 coordinates of the sphere $S^7$.
 
The variables parametrizing the sphere $S^7$ 
correspond to `helicity'
  degrees of freedom of the quantum states of the massless $D=10$
superparticle.
\bigskip

It is worth mentioning that the appearance of extra compact
dimensions in the momentum spaces of the superparticle models
considered 
above is
related to the well known  fact that in $D=3,4,6$ and 10
the  commuting spinors (twistors) with $n=2(D-2)=2,4,8$ and 16
components
parametrize, modulo scale transformations,
   $S^1$, $S^3$, $S^7$ and $S^{15}$ spheres, respectively. 
These spheres are Hopf fibrations (fiber bundles) which are associated
with the division algebras {\bf R,~C,~H} and {\bf O}.
Their bases are 
the spheres $S^1$, $S^2$, $S^4$ and $S^8$, and the fibers are $Z^2$, 
$S^1=U(1)$, $S^3 =SU(2)$
and $S^7$, respectively. The base spheres are
parametrized (up  to a scaling factor 
which, due to the Cartan-Penrose representation, 
is identified with the square of 
spinor components) 
by the 
light--like vectors (massless particle momenta) in $D=3,4,6$ and $10$, 
respectively.
We see that the  fibers are extra ``momentum dimensions" which we have
in our 
models with 
central charges (at $a\not = 0$). This is the geometrical ground for 
the appearance of  $S^{1}$ in $D=4$ and  $S^7$ in $D=10$.

\section{Conclusion and discussion}

We have performed the detailed analysis and quantization of the
massless superparticle model with tensorial central charges associated
with
twistor--like commuting spinors in space--times of dimension $D=3,4,6$
and 10.
The physical phase space  degrees of
freedom of this model have a natural description in terms of
supertwistors
which form a fundamental representation of a corresponding maximal
  supergroup
of conformal type underlying the dynamics of the superparticle.

A peculiarity of the $a=1$ model is that it possesses
$n=2^{[{D\over 2}]}-1$ $\kappa$--symmetries, while the standard massless
superparticles have $n=2^{[{D\over 2}]-1}$  $\kappa$--symmetries. 
The presence of such a large number of $\kappa$--symmetries in the 
$a=1$ models
means that the superparticle breaks only one of the $2^{[{D\over 2}]}$ 
supersymmetries of the target space vacuum. This results in very short
 two--component
supermultiplets describing the quantum states of the $a=1$
superparticle,
since the corresponding target space superfields depend only on one
Grassmann coordinate. The existence of these short Lorentz--covariant
superfields is made possible because the target superspace has been
enlarged by commuting spinor coordinates, whose role is in singling out
a `small' covariant subsuperspace in the extended target superspace.
Let us compare this situation with well known cases.

In the case of the ordinary massless superparticle in N=1, D=4
superspace the
quantum states of the superparticle are described by a chiral scalar 
superfield \cite{n1}.  The
chirality constraint is a consequence of first--class fermionic constraints 
 generating $\kappa$--symmetries.
 Consequently the 
 chiral superfield effectively depends on only two Grassmann 
coordinates, which reflects the fact that the ordinary superparticle
preserves half (i.e. two out of four) supersymmetries of 
$N=1$, $D=4$ superspace.

In the case of an $N=2$, $D=4$ superparticle 
in harmonic superspace \cite{n2}
which also breaks half of the target space supersymmetries, 
$SU(2)$--harmonic variables allow one to
pick a harmonic analytic subsuperspace out of the general 
$N=2$, $D=4$ superspace \cite{harm}, and  
quantum states of the superparticle are described by analytic
superfields
which depend on four Grassmann coordinates singled out from 
the original eight Grassmann coordinates 
by the use of the harmonic variables.

In the analogous way, in the case of the generalized superparticle model
\p{ac(a)}, \p{actionD}
at $a=1$, when only one target space supersymmetry is broken, one finds
a Lorentz covariant subsuperspace of the target superspace,
 which has only one Grassmann direction parametrized
by the Lorentz scalar $\theta^\a\lambda_\a$. The ``short'' superfields
\p{wf4Ds3}, \p{Psisol}, 
which exist only due to the presence of the auxiliary spinor 
 variable, 
 describe the quantum states of the generalized 
superparticle.

We have shown that in contrast to standard superparticles the 
considered model
possesses additional compact phase--space variables which
describe helicity degrees of freedom of the superparticle and which
upon quantization parametrize infinite tower of free states with
arbitrary
(half)integer helicities. Due to this property it would be  
interesting
to consider the possibility of treating 
our generalized superparticle
model as a classical mechanics counterpart 
of the theory of higher--spin fields
developed by M. Vasiliev \cite{Vasiliev}. Since the nontrivially
interacting
higher spin fields should live in a space--time of (anti)-de--Sitter 
geometry
a natural generalization of the results of this paper would be
to consider a superparticle model on supergroup manifolds
describing isometries of corresponding AdS superspaces. For 
 $D=4$ 
 the supergroup $OSp(1|4)$ is the isometry of 
a $D=4$ AdS superspace ${OSp(1|4)}\over{SO(1,3)}$ 
which in addition to 4 bosonic directions
has 4 Grassmann fermionic directions. Six bosonic coordinates
corresponding
to the group $SO(1,3)$ (which extends 
the coset superspace ${OSp(1|4)}\over{SO(1,3)}$ 
to the supergroup manifold $OSp(1|4)$) are a non--Abelian 
generalization of the central charge  coordinates of the $D=4$ model
considered
above. It appears that our model with central charges 
 can  be regarded as an appropriate
truncation of the  $OSp(1|4)$ model. Work in this direction is now in
progress.

\bigskip 
We should also remark that
tensorial central charges are usually associated with 
brane charges, which are topological and take discrete 
(quantum) values. In contrast, in the superparticle models considered in
this paper the central charges 
take continuous values and parametrize compact manifolds, while their
Fourier conjugate coordinates are quantized.

\bigskip
{\bf Acknowledgements}. The authors would like to thank J. de Azcarraga, 
K. Landsteiner, C. Preitschopf and  M. Tonin 
for interest to this work and valuable discussions.
Work of I.B. and D.S. was partially supported by 
the INTAS Grants  N96--308 and N 93--493--ext.


\newpage

\noindent
\section*{APPENDIX} 
\setcounter{equation}{0}

\subsection*{A1. Quantization of one fermionic degree of freedom by a 
`half-conversion' prescription}

Here we shall present a method of quantizing a single
fermionic variable alternative to that used in Subsection 3.1.4, but
which leads to the same spectrum of quantum states.

Let us convert the second--class constraint \p{S1} into the first--class
constraint by  introducing one more Clifford-like variable 
$\xi^\prime$
\begin{equation}\plabel{Pbrxi'}
\{ \xi^\prime , \xi^\prime \}_P = - {i \over 2}. \qquad 
\end{equation}
Using $\xi^\prime$ we replace \p{S1} by the first class
constraint
\begin{equation}\plabel{Azeta}
{\cal A} \equiv  \xi -i \pi_{(\xi)} + 2 \xi^\prime
=0, \qquad \{ {\cal A} , {\cal A} \}_P= 0,
\end{equation}
where instead of $\chi_1$ of Eq. \p{S1} we have introduced ${\xi
=2\chi_1}$.
\bigskip


Let us quantize the model using the coordinate representation for the
original Grassmann variable $\xi$ 
$$
\hat{\xi}=\xi, \qquad \hat{\pi}_\xi = i {\partial\over \partial \xi }
$$
and a real $2 \times 2$ matrix representation 
\begin{equation}\plabel{xitau}
\hat{\xi^\prime}=  {1 \over 2} \tau = {1 \over 2} \left(\matrix{ * & *
\cr 
                                                    * & * \cr}\right) 
\qquad 
\tau^2=I 
\end{equation} 
for the new Clifford algebra valued variable $\hat{\xi^\prime}$ 
$$
\hat{\xi^\prime}^2 = 1/4 . 
$$

Then the wave function is regarded as a column 
 \begin{equation}\plabel{Psicol}
\Psi_a=  \left(\matrix{ \phi (\xi) \cr  
                         \psi (\xi) \cr}\right)  
\end{equation}
and the quantum counterpart of the first class constraint 
\p{Azeta} 
\begin{equation}\plabel{hatA}
\hat{{\cal A}} \equiv  
(\xi - {\partial \over \partial \xi }) 
 \tau^\prime  + \tau  \qquad 
\end{equation}
should be imposed on the wave function 
\begin{equation}\plabel{APsi=0}
\hat{{\cal A}}_{ab}\Psi_b \equiv  
[(\xi - {\partial \over \partial \xi })~  
 \tau^\prime_{ab}  + \tau_{ab}]\Phi_b (\xi) =  0 . \qquad 
\end{equation}

In \p{hatA} and \p{APsi=0} the second $2 \times 2$
matrix 
\begin{equation}\plabel{tau'}
\tau^\prime_{ab} =  \left(\matrix{ * & * \cr 
                                                    * & * \cr}\right), 
\qquad (\tau^\prime)^2=I
\end{equation}
was introduced. 
It is required to ensure the anticommutativity of the Grassmann and 
Clifford part of the first class constraint \p{APsi=0}. 
Indeed, 
let us calculate  the square of  the 
quantum constraint \p{hatA}
\begin{equation}\plabel{hatA20}
\hat{{\cal A}}^2 
 \equiv {1 \over 2}\{ \hat{{\cal A}}, \hat{{\cal A}}\} = 
\tau ^2 + (\xi - {\partial \over \partial \xi })^2 (\tau^\prime)^2 + 
 (\xi - {\partial \over \partial \xi }) 
\{ \tau^\prime , \tau\} . \qquad 
\end{equation}
Since
\begin{equation}\plabel{op2}
(\xi - {\partial \over \partial \xi })^2= 
{1 \over 2} \{ (\xi - {\partial \over \partial \xi }),  
(\xi - {\partial \over \partial \xi }) \} = -1 
\end{equation}
and 
$$
\tau^2 = I = (\tau^\prime)^2, 
$$
one easily finds that the first two terms in \p{hatA20} cancel and
arrives at
\begin{equation}\plabel{hatA2}
\hat{{\cal A}}^2 =
  (\xi - {\partial \over \partial \xi }) 
{1 \over 2} \{ \tau^\prime , \tau\} . \qquad 
\end{equation}
The last input vanishes if and only if 
$\{ \tau^\prime , \tau\}=0$. 
This result can not be reached if one chose $ \tau^\prime$
to be the unit matrix. $ \tau$ and $ \tau^\prime$ can be chosen to be
two 
Pauli matrices. 

Let us stress  that the necessity to introduce the second matrix 
$ \tau^\prime$ is a peculiarity of the quantization of the odd number
of  Clifford variables. $ \tau^\prime$ can be the unit matrix in the
case of 
even number of Clifford variables 
(see e.g. \cite{even} and references therein).

To fix the representation for the matrices $\tau$ and $\tau^\prime$ one
has to 
note that the conservation of the Grassmann parity in the form of the
first 
class constraint \p{APsi=0} requires that
\begin{itemize}
\item The components $\phi (\xi)$ and $ \psi (\xi)$ 
of 
\p{xitau} must have {\sl different Grassmann parity}. For instance, if 
we choose $\phi (\xi)$ to be bosonic superfield  then $\psi (\xi)$
is fermionic.
\item If the diagonal representation is chosen for one  of the
matrices, 
say  
$\tau^\prime$, then another matrix $\tau$ is antidiagonal.   
\end{itemize}

Taking these in mind we choose 
\begin{equation}\plabel{taurep}
\tau^\prime = \tau_3 \equiv \left(\matrix{ 1 & 0 \cr 
                                                    0 & -1 \cr}\right), 
\qquad 
\tau = \tau_1 \equiv  \left(\matrix{ 0 & 1 \cr 
                                                    1 & 0 \cr}\right). 
\qquad 
\end{equation}
Then the quantum constraints 
\p{APsi=0} acquire the form 
\begin{equation}\plabel{hatAPsi=0}
\hat{{\cal A}}_{ab}\Psi_b \equiv  
\left(\matrix{ (\xi - {\partial \over \partial \xi })~  & 1 \cr 
                1 & - (\xi - {\partial \over \partial \xi })
                          \cr}\right) 
\left(\matrix{ \phi (\xi) &\cr 
                         \psi (\xi)& \cr}\right)_b = 
\left(\matrix{ (\xi - {\partial \over \partial \xi }) \phi (\xi)
+ \psi (\xi) \cr 
                 \phi (\xi) - (\xi - {\partial \over \partial \xi
})
                         \psi (\xi) \cr}\right) = 0  , \qquad 
\end{equation}
which splits into two equations 
\begin{equation}\plabel{eqs}
(\xi - {\partial \over \partial \xi }) \phi (\xi)
 =-  \psi (\xi) , \qquad
  \end{equation}
$$
  \phi (\xi) = (\xi - {\partial \over \partial \xi })
                         \psi (\xi) 
$$
Using \p{op2}, we notice that the second 
equation is a consequence of the first one. 
The first equation 
\begin{equation}\plabel{eq}
(\xi - {\partial \over \partial \xi }) \phi (\xi)
 =-  \psi (\xi) \qquad
  \end{equation}
expresses the fermionic superfield through the bosonic one. 
I.e. if we write $\phi (\xi)$ in components
$$
\phi (\xi) = \phi_0 + i \xi \psi_1,  
$$
then from eq. \p{eq} it follows that
$$
\psi (\xi) = i \psi_1 -  \xi \phi_0. 
$$

Thus, we can represent the spectrum of states carrying 
one Clifford degree of freedom
by one (either bosonic or fermionic) superfield $\phi (\xi)$
depending on the single {\sl Grassmann} variable $\xi$
$$
\xi^* = \xi, \qquad \xi^2 = 0. \qquad 
$$ 

This result is in accordance with that of Subsection 3.1.4 
(see also \cite{Sorokin87}), 
and both methods of quantizing a single fermionic variable 
result in the same field content of quantum states 
(one boson and one fermion). 

\subsection*{A2. Quantization of the high-dimensional model with the 
use of conversion}

Here we present some details of getting the wave function \p{Psisol}
from the converted system of constraints 
\p{tPhi}, \p{tD} and \p{tPl}
describing the high-dimensional generalization of 
the first--quantized $a=1$ model.

Let us choose the (super)coordinate 
representation for supercoordinates and bosonic spinors 
\begin{equation}\plabel{hatP}
\hat{P}_{\a\b} = - i 
{\partial \over \partial X^{\a\b}}, \qquad 
\hat{P}^{\a}_{(\tilde{\l}) }= - i 
{\partial \over \partial \tilde{\l}^{\a}}, 
\end{equation}
\begin{equation}\plabel{hatpi}
\hat{\pi}_{\a} = i 
{\partial \over \partial \Theta^{\a}} 
\end{equation}
and use the $2\times 2$ matrix representation 
 \begin{equation}\plabel{xitauD}
\hat{\xi}=  {1 \over 2} \tau_2 = 
\left(\matrix{ 0 & 1 \cr 
               1 & 0 \cr}\right),
\quad
\{ \hat{\xi}, \hat{\xi}\} = {1 \over 2}.
\end{equation}
for the Clifford variable $\xi$.

Then the the wave function is  a column
 \begin{equation}\plabel{wfD}
\Psi_a =  \Psi_a (X^{\a\b}, \tilde{\l}_\b) 
=
\left(\matrix{  \phi (X^{\a\b}, \tilde{\l}_\b, \tilde{\rho}_\g)  \cr 
                \psi (X^{\a\b}, \tilde{\l}_\b) \cr}\right)
\end{equation}
with the elements carrying opposite Grassmann parity 
(e.g. $\phi$ is bosonic and $\psi$ is fermionic) 
and the 
quantum first class constraints \p{tPhi}, \p{tD}, \p{tPl}
should be taken in the form
\begin{equation}\plabel{qtildePhi}
\hat{\tilde{\Phi}}_{\a\b} = 
- i (\partial_{\b\a}- i \tilde{\l}_\a \tilde{\l}_\b)~I 
\end{equation}
\begin{equation}\plabel{qtildeD}
\hat{\tilde{D}}_\a = \hat{{D}}_\a + \tilde{\l}_\a \hat{\xi} = 
i (\partial_\a - i \Theta^\b \partial_{\b\a}) \tau_3 
+ \tilde{\l}_\a \tau_2).  
\end{equation}

The incorporation  of the $\tau_3 $ matrix 
is necessary to provide the properties of the first class constraints 
to form the closed algebra
\begin{equation}\plabel{qDbr}
\{ \hat{\tilde{D}}_\a, \hat{\tilde{D}}_\b \} = 
- 2i \hat{\tilde{\Phi}}_{\a\b}. 
\end{equation}
This is a peculiarity of the quantization of the models with  odd number 
of phase space Grassmann variables 
(see  section A1. of this Appendix). 

\bigskip 

The further steps of the quantization procedure exactly repeat
the steps of the $D=4$ case (see Section 4).  
The  wave function describing the spectrum of the quantum states is
\begin{equation}\plabel{Psisolc}
 \tilde{\Psi_a}= e^{{i \over 2} \tilde{\l}_\a\tilde{\l}_\b X^{\a \b}}
\left(\matrix{ \Psi (\tilde{\l}_a, \Th\l )  \cr 
-i \left(\partial_\chi + \chi\right) \Psi (\tilde{\l}_a, \Th\l ).
\cr}\right)\end{equation}
As the second element in the column is expressed through the first one, 
we can describe the spectrum of the quantum states by the single superfield
\p{Psisol} depending on  bosonic  
$\tilde{\l}_\a$ and fermionic $\chi= (\Th\l)$.
 

\end{document}